\documentclass[aps,twocolumn,superscriptaddress,floatfix,nofootinbib,showpacs,altaffilletter]{revtex4-2}

\usepackage{epsfig,graphicx,amssymb,amsmath,wasysym,graphicx}

\usepackage{amsmath}
\usepackage{amsfonts}
\usepackage{amssymb}
\usepackage{graphicx}
\usepackage{bm}
\usepackage{subfigure}
\usepackage{url}
\usepackage[hyperindex]{hyperref}
\usepackage{color}
\usepackage[ddmmyy,24hr]{datetime}
\usepackage{bigdelim}
\usepackage{booktabs}
\usepackage{dcolumn}
\usepackage{multirow}
\usepackage{subfigure}
\usepackage{cancel}
\usepackage{stackrel}
\usepackage{paralist}
\usepackage{xspace}
\usepackage{slashed}
\usepackage{cancel}
\usepackage{todonotes}
\usepackage{enumerate}
\usepackage{float}
\usepackage{fullpage}
\usepackage{ulem}
\usepackage{physics}

\usepackage{lineno}

\newcommand{\nua}[1]{\ensuremath{\rlap{\kern-2.5pt\ensuremath{\overset{\scriptscriptstyle(-)}{\phantom{\nu}}}}{\ensuremath{{\nu}_{#1}}}}}

\usepackage{enumitem}

\newcommand{\cenns}{CE$\nu$NS\xspace}

\newcommand{\be}{\begin{equation}}
\newcommand{\ee}{\end{equation}}
\newcommand{\ba}{\begin{array}}
\newcommand{\ea}{\end{array}}

\begin{document}

\title{On the impact of the Migdal effect in reactor CE$\nu$NS experiments}

\author{M. Atzori Corona}
\email{mattia.atzori.corona@ca.infn.it}
\affiliation{Dipartimento di Fisica, Universit\`{a} degli Studi di Cagliari,
	Complesso Universitario di Monserrato - S.P. per Sestu Km 0.700,
	09042 Monserrato (Cagliari), Italy}
\affiliation{Istituto Nazionale di Fisica Nucleare (INFN), Sezione di Cagliari,
	Complesso Universitario di Monserrato - S.P. per Sestu Km 0.700,
	09042 Monserrato (Cagliari), Italy}

\author{M. Cadeddu}
\email{matteo.cadeddu@ca.infn.it}
\affiliation{Istituto Nazionale di Fisica Nucleare (INFN), Sezione di Cagliari,
	Complesso Universitario di Monserrato - S.P. per Sestu Km 0.700,
	09042 Monserrato (Cagliari), Italy}

\author{N. Cargioli}
\email{nicola.cargioli@ca.infn.it}
\affiliation{Dipartimento di Fisica, Universit\`{a} degli Studi di Cagliari,
	Complesso Universitario di Monserrato - S.P. per Sestu Km 0.700,
	09042 Monserrato (Cagliari), Italy}
\affiliation{Istituto Nazionale di Fisica Nucleare (INFN), Sezione di Cagliari,
	Complesso Universitario di Monserrato - S.P. per Sestu Km 0.700,
	09042 Monserrato (Cagliari), Italy}

\author{F. Dordei}
\email{francesca.dordei@cern.ch}
\affiliation{Istituto Nazionale di Fisica Nucleare (INFN), Sezione di Cagliari,
	Complesso Universitario di Monserrato - S.P. per Sestu Km 0.700,
	09042 Monserrato (Cagliari), Italy}

\author{C. Giunti}
\email{carlo.giunti@to.infn.it}
\affiliation{Istituto Nazionale di Fisica Nucleare (INFN), Sezione di Torino, Via P. Giuria 1, I--10125 Torino, Italy}


\begin{abstract}

The search for coherent elastic neutrino nucleus scattering (CE$\nu$NS) using reactor antineutrinos represents a formidable experimental challenge, recently boosted by the observation of such a process at the Dresden-II reactor site using a germanium detector. This observation relies on an unexpected enhancement at low energies of the measured quenching factor with respect to the theoretical Lindhard model prediction, which implies an extra observable ionization signal produced after the nuclear recoil. A possible explanation for this additional contribution could be provided by the so-called Migdal effect, which however has never been observed. Here, we study in detail the impact of the Migdal contribution to the standard CE$\nu$NS signal calculated with the Lindhard quenching factor, finding that the former is completely negligible for observed energies below $\sim 0.3\,\mathrm{keV}$ where the signal is detectable, and thus unable to provide any contribution to CE$\nu$NS searches in this energy regime. 
To this purpose, we compare different formalisms used to describe the Migdal effect that intriguingly show a perfect agreement, making our findings robust.   

\end{abstract}

\maketitle  

\section{Introduction}

Coherent elastic neutrino-nucleus scattering (\cenns) is a pure weak neutral current low-energy process predicted by Freedman in 1973~\cite{Freedman:1973yd} where the neutrino interacts with the nucleus as a whole. Namely, all the nucleons within the nucleus respond coherently to the neutrino interaction, leading to a higher cross section than other low-energy processes involving neutrinos. The remarkable observations of \cenns from the COHERENT Collaboration~\cite{COHERENT:2017ipa,COHERENT:2020iec,COHERENT:2021xmm} have opened a new era to test our knowledge of the Standard Model (SM) of particle physics and has posed an exciting technological challenge to develop innovative detectors capable of spotting the extremely tiny nuclear recoils produced as a single outcome of the interaction~\cite{NUCLEUS:2019igx,Strauss:2017cuu,Akimov:2022oyb,Baxter:2019mcx,CONUS:2020skt,CONNIE:2016nav,Billard:2016giu,nGeN:2022uje,Colaresi:2021kus,Colaresi:2022obx,CONNIE:2021ggh,MINER:2016igy,Su:2023klh, Colas:2021pxr, Akimov:2022xvr, Wong:2015kgl}. These experiments play a crucial role in advancing our knowledge of neutrino interactions and for their implications for fundamental physics~\cite{Cadeddu:2020nbr, Cadeddu:2018izq, DeRomeri:2022twg, Giunti:2019xpr, Lindner:2016wff, Billard:2018jnl, Cadeddu:2017etk, Cadeddu:2021ijh, Cadeddu:2018dux, Cadeddu:2020lky, AristizabalSierra:2018eqm, Cadeddu:2019eta, AtzoriCorona:2022moj, Denton:2018xmq, Denton:2020hop, Co:2020gwl}.

\cenns requires a high neutrino flux to produce signal events above experimental backgrounds. Among the different neutrino sources available, in this study, we focus on \cenns produced from reactor antineutrinos and observed with germanium detectors.  
There are three main germanium detectors currently operating, namely NCC-1701~\cite{Colaresi:2021kus, Colaresi:2022obx} (also referred to as Dresden-II), CONUS~\cite{Bonet:2020ntx} and $\nu$GEN~\cite{nGeN:2022uje}, located 10.39~m, 17.1 m and 11 m away from $2.96\;\rm{GW}_{\rm{th}}$, $3.9\;\rm{GW}_{\rm{th}}$, $3.1\;\rm{GW}_{\rm{th}}$ commercial reactors, respectively.
Interestingly, the Ricochet experiment at the ILL site, 8.8~m away from the core of the $58.3\; \rm{MW}_{\rm{th}}$ research nuclear reactor, is also aiming to measure \cenns down to the sub-100~eV nuclear energy recoil regime~\cite{Ricochet:2022pzj}.

In particular, the recent first observation of \cenns at the Dresden-II reactor~\cite{Colaresi:2022obx} has gained a lot of attention due to the broad impact of such a result on current and future \cenns searches, and the physics that can be extracted within the SM and beyond~\cite{AtzoriCorona:2022qrf,Coloma:2022avw,Majumdar:2022nby, Denton:2022nol}.
As a matter of fact, this measurement is highly affected by the knowledge of the germanium quenching factor (QF) at low nuclear recoil energies. The QF quantifies the reduction of the ionization yield produced by a nuclear recoil with respect to an electron recoil of the same energy.
Indeed, the \cenns observation by Dresden-II depends crucially on the two 
new QF measurements reported in Ref.~\cite{Collar:2021fcl}.  
They have been obtained from photo-neutron source measurements, so-called YBe, and from iron-filtered monochromatic neutrons, so-called Fef~\cite{Collar:2021fcl}. However, these two QF determinations are in contrast with and significantly higher than the standard Lindhard prediction with the parameter $k=0.157$~\cite{Lindhard_theo} and other independent experimental measurements (see e.g. Ref.~\cite{SuperCDMS:2022nlc} for a recent measurement and the summary plot in Fig.~1). Moreover, CONUS data disfavours quenching parameters above $k = 0.27$~\cite{CONUS:2020skt} and a recent low-energy determination of the QF finds a good agreement with the Lindhard theory with a parameter $k=0.162\pm0.004$~(stat+sys)~\cite{Bonhomme:2022lcz}.

Among the possible solutions to explain the increase of the QF at very low energies, it has been proposed that the Lindhard model might be not sufficient to encode the full behavior of the QF in the low-energy regime. In particular, it has been suggested~\cite{Collar:2021fcl} that the Migdal effect may play a crucial role. This yet unobserved process\footnote{On the same day our manuscript was made public on arXiv, another one appeared on this topic~\cite{Xu:2023wev}. In this work, the authors reported the first direct search for the Migdal effect in liquid xenon using nuclear recoils produced by tagged neutron scatters. If the suppression of the Migdal effect with respect to the theoretical prediction found in Ref.~\cite{Xu:2023wev} happens also in germanium, the explanation of the high QF found in Ref.~\cite{Collar:2021fcl} through the Migdal effect is even less likely than that found in this work.} has been first proposed in 1941~\cite{Migdal} and recently taken into serious consideration in the context of dark matter searches~\cite{Ibe:2017yqa,Essig:2019xkx,GrillidiCortona:2020owp,Liu:2020pat,DarkSide:2022dhx,XENON:2019zpr,SuperCDMS:2023sql,Bell:2021zkr} and neutrino physics~\cite{Bell:2019egg,Bell:2021ihi}.
The Midgal effect might happen after a nuclear recoil is induced by a neutral particle, i.e., a neutron, a neutrino, or dark matter, due to the displacement between the recoiling nucleus and the electronic cloud.  Indeed, the atomic electrons do not immediately follow the motion of the recoiling nucleus. Thus, in order to restore equilibrium, extra ionization can be injected into the detector. 
In Refs.~\cite{Liao:2021yog, Liao:2022hno} there have been attempts to include the Migdal effect by means of some phenomenological parameters originally introduced to go beyond the standard Lindhard theory~\cite{Sorensen:2014sla}.
However, the approach exploited in these studies neglects the microscopic physics of the Migdal process that can instead be accounted for using existing formalisms.

The goal of this paper is to perform a detailed characterization of the Migdal effect in a germanium detector searching for \cenns at a reactor site. To do so, we use different approaches to describe the microscopic physics of the phenomenon to quantify the robustness of our findings. Finally, we discuss to which extent the Migdal effect plays a role in the observation of \cenns at the Dresden-II reactor and thus, whether it can be claimed as an explanation of the anomalous enhancement of the QF at low energies.

\section{\cenns}

The differential \cenns cross section for a neutrino of flavor $\mathcal{\ell}$ that scatters off a nucleus $\mathcal{N}$ with $Z$ protons and $N$ neutrons is given by~\cite{Freedman:1973yd,PhysRevD.30.2295,Barranco:2005yy,AtzoriCorona:2023ktl,Cadeddu:2023tkp}
\begin{align}\nonumber
\dfrac{d\sigma_{\nu_{\ell}\text{-}\mathcal{N}}}{d T_\mathrm{nr}}
	& = 
	\dfrac{G_{\text{F}}^2 M}{\pi}
	\left( 1 - \dfrac{M T_\mathrm{nr}}{2 E_\nu^2} \right)\times\\	&\left[g_{V}^{p}\left(\nu_{\ell}\right) Z F_{Z}\left(|\vec{q}|^{2}\right)+g_{V}^{n} N F_{N}\left(|\vec{q}|^{2}\right)\right]^{2},
\label{eq:SMcevns}
\end{align}

where $T_{\rm{nr}}$ is the nuclear recoil energy,  $E_\nu$ is the neutrino energy, and the term inside the square brackets is the so-called weak charge of the nucleus $\mathcal{Q}_W$.
In Eq.~(\ref{eq:SMcevns}) $G_{\text{F}}$ is the Fermi constant and $M$ the nuclear mass.
For Ge ($Z$=32) we use $N$=(38, 40, 41, 42, 44) corresponding to the natural abundances of 0.2057 ($\mathrm{{}^{70}Ge}$), 0.2745 ($\mathrm{{}^{72}Ge}$), 0.0775 ($\mathrm{{}^{74}Ge}$), 0.3650 ($\mathrm{{}^{74}Ge}$), 0.0773 ($\mathrm{{}^{76}Ge}$)~\cite{BerglundWieser+2011+397+410}.
$F_{Z}\left(|\vec{q}|^{2}\right)$ and $F_{N}\left(|\vec{q}|^{2}\right)$ are, respectively, the proton and neutron form factors that parameterize the loss of coherence for increasing values of the momentum transfer $|\vec{q}|$. 
In the low-energy regime that characterizes reactor neutrino experiments, the form factors are practically equal to one, so basically the result does not depend on the parametrization used as well as on the values of the proton and neutron distribution radii. The neutrino-nucleon couplings are computed using the radiative corrections as described in Refs.~\cite{ParticleDataGroup:2022pth,AtzoriCorona:2023ktl}
that yield to $g_V^p(\nu_e)=0.0382$ and $g_V^n=-0.5117$.

For a \cenns experiment located at a reactor site, only the $\bar{\nu}_e$ flavor contributes.
The antineutrino spectrum $d N_{\bar{\nu}_e}/d E_\nu$ is obtained by combining the expected spectra for $E_\nu>2$~MeV from Ref.~\cite{Huber:2011wv,Mueller:2011nm} with the low energy part determined in Ref.~\cite{PhysRevD.39.3378}. 
This parameterization is referred to as HMVE, and it has been shown~\cite{AtzoriCorona:2022qrf} that the usage of other parameterizations~\cite{Estienne:2019ujo,Kopeikin:1999tc,Kopeikin:2012zz} does not significantly affect the results of the Dresden-II analysis.

When an isolated nuclear recoil occurs, the electron-equivalent (ee) nuclear recoil energy observed in the detector, $E_{\rm{det}}$, is expressed as
\begin{equation}
    E_{\rm{det}}=f_Q(T_{\rm{nr}})T_{\rm{nr}},
    \label{eq:quenching}
\end{equation}
where $f_Q(T_{\rm{nr}})$ is the energy-dependent parameterization of the QF.
The theoretical \cenns event rate is therefore
\begin{align}\nonumber
    \frac{dR}{dE_{\rm{det}}}=&N_T(\textrm{Ge})\int_{E_{\nu}^{\rm{min}}}^{E_\nu^{\rm{max}}}dE_\nu\frac{d N_{\bar{\nu}_e}}{d E_\nu}\dfrac{d\sigma_{\bar{\nu}_e\text{-}\mathcal{N}}}{d T_\mathrm{nr}}\times\\
    &\left(f_Q+T_{\rm{{nr}}}\frac{df_Q}{dT_{\rm{nr}}}\right)^{-1} ,
    \label{eq:ratecevns}
\end{align}
where $N_T(\textrm{Ge})$ is the number of target atoms in the detector, $E_\nu^{\text{min}}(T_{\text{nr}}) \simeq \sqrt{MT_\text{nr}/2}$ and $E_\nu^{\text{max}} \simeq 10$~MeV.
The last term in Eq.~(\ref{eq:ratecevns}) is $dT_{\rm{nr}}/dE_{\rm{det}}$, which is needed to express the rate in terms of the electron-equivalent nuclear recoil energy defined in Eq.~(\ref{eq:quenching}).

\section{The Migdal effect}\label{sec:migdal}

Despite the very intense $\bar{\nu}_e$ flux, the search for \cenns from reactor neutrinos is very challenging due to the tiny signal produced and the still-under-debate behavior of the quenching factor at low energies. In this scenario, it is crucial to characterize the impact of the Migdal effect to prevent potential misinterpretations of experimental data. We first calculate the Migdal rate using the formalism of Ibe \textit{et al.}~\cite{Ibe:2017yqa}, which considers the target as composed of isolated atoms. Moreover, this formalism relies on the dipole approximation that allows one to write the Migdal transition matrix element, $M_{fi}$, in the form
\begin{align}\nonumber
    M_{fi}&=\langle \psi_f|e^{-i m_e\vec{v}\cdot \sum_{i=1}^{Z}\vec{r}_i}|\psi_i \rangle\simeq\\ \nonumber
    &\simeq -im_e\vec{v}\cdot \langle\psi_f|\sum_{i=1}^{Z}\vec{r}_i|\psi_i \rangle \\
    &\equiv -im_e\vec{v}\cdot\vec{D}_{fi},\label{eq:dipole}
\end{align}
where $\vec{r}_i$ is the position operator of the $Z$ electrons, $\vec{D}_{fi}$ is the dipole matrix element, $m_e$ is the electron mass,  $\vec{v}$ is the nuclear recoil velocity, while $\psi_f$ and $\psi_i$ are the wavefunctions of the final and the initial atomic states in the nucleus rest frame. 
The final state wavefunctions are boosted to the rest frame of the recoiling nucleus by a Galilean transformation and are computed using the Dirac-Hartree-Fock method.

Under these assumptions, the differential cross section for the Migdal effect can be written as
\begin{align}\nonumber
\left(\dfrac{d\sigma_{\bar{\nu}_e\text{-}\mathcal{N}}}{d T_\mathrm{nr}}\right)_{\rm{Migdal}}^{{\rm{Ibe}}\;et\;al.}
	 = &
	\dfrac{G_{\text{F}}^2 M}{\pi}
	\left( 1 - \dfrac{M T_\mathrm{nr}}{2 E_\nu^2} \right)\mathcal{Q}_W^2\times
    \\
     & \left|Z_{\rm{ion}}(q_e)\right|^2,
\label{eq:MigdalIbe}
\end{align}
where $\left|Z_{\rm{ion}}(q_e)\right|$ is the ionization rate of an individual electron in the target with momentum $q_e$. It is defined as
\begin{equation}
\left|Z_{\rm{ion}}(q_e)\right|^2 = \frac{1}{2\pi}\sum_{n,\ell}\int dT_e\frac{d}{dT_e}p^c_{q_e}(n\ell\rightarrow T_e),
\end{equation}
where $p^c_{q_e}(n\ell\rightarrow T_e)$ are the ionization probabilities for an atomic electron with quantum numbers $n$ and $\ell$ that is ionized with a final energy $T_e$. It should be noticed that very similar results are expected if one relies on the probabilities calculated in Ref.~\cite{Cox:2022ekg} with an independent approach. Indeed, the authors have shown a very good agreement with the results obtained with the Ibe \textit{et al.} formalism that is also used in this work, demonstrating that the addition of semi-inclusive ionization probabilities is not significant at low recoil energies for atomic germanium.

The double differential cross section for the $n\ell$ state contribution as a function of both the electron and the nuclear recoil energy is hence
\begin{align}\nonumber
\left(\dfrac{d^2\sigma_{\bar{\nu}_e\text{-}\mathcal{N}}}{d T_\mathrm{nr}dT_e}\right)_{n\ell}^{{\rm{Ibe}}\;et\;al.}
	 =& 
	\dfrac{G_{\text{F}}^2 M}{\pi}
	\left( 1 - \dfrac{M T_\mathrm{nr}}{2 E_\nu^2} \right)\mathcal{Q}_W^2\times\\
 &\frac{1}{2\pi}\, 
 \frac{d}{dT_e}p^c_{q_e}(n\ell\rightarrow T_e).
\label{eq:doublecsecMigdalIbe}
\end{align}

If the nuclear recoil is followed by a Migdal emission, the total energy deposit of the event in the detector is 
\begin{equation}
    E_{\rm{det}}=f_QT_{\rm{nr}}+T_e+E_{n\ell},
\end{equation}
where the first term is the nuclear recoil energy deposit, while $T_e$ and $E_{n\ell}$ account for the extra energy injected in the detector, $E_{n\ell}$ being the atomic de-excitation energy for Ge~\cite{Ibe:2017yqa}.
 We evaluate the theoretical event rate as a function of the detected energy, which is given by  
\begin{align}\nonumber
    \left(\frac{dR}{dE_{\rm{det}}}\right)_{\textrm{Migdal}}^{{\rm{Ibe}}\;et\;al.}&=N_T(\textrm{Ge})\sum_{n,\ell}\int_{E_{\nu}^{\rm{min}}}^{E_\nu^{\rm{max}}}dE_\nu\frac{d N\bar{\nu}_e}{dE_\nu}\times
    \\ \nonumber   
    &\int dT_{e} \int_{T_\textrm{nr}^{\rm{min}}}^{T_{\rm{nr}}^{\rm{max}}} dT_{\rm{nr}}\left(\dfrac{d^2\sigma_{\bar{\nu}_e\text{-}\mathcal{N}}}{d T_\mathrm{nr}dT_e}\right)_{n\ell}^{{\rm{Ibe\; } }et\;al.} \times\\
 & \delta(E_{\rm{det}}-f_Q T_{\rm{nr}}-T_e-E_{n\ell}),
    \label{eq:rateMigdal}
\end{align}
where we have imposed energy conservation using the Dirac $\delta$ and $T_{\rm{nr}}$ is now constrained within the values ${T_{\rm{nr}}^{\rm{min}}}$ and ${T_{\rm{nr}}^{\rm{max}}}$ given by~\cite{Bell:2019egg}

\begin{equation}
\frac{\left(T_e + E_{n\ell}\right)^2}{2M} \leq T_{\rm{nr}} \leq \frac{\left(2E_\nu - (T_e + E_{n\ell})\right)^2}{2(M + 2E_\nu)}.
\label{eq:range}
\end{equation}
The rate in Eq.~(\ref{eq:rateMigdal}) represents the Migdal contribution summed over all the possible $n\ell$ atomic states. 
The total predicted event rate is thus given by the sum of Eq.~(\ref{eq:ratecevns}) and Eq.~(\ref{eq:rateMigdal}).

\section{Migdal Photo Absorption}

The formalism described so far to compute the dipole matrix element for the Migdal rate relies on the assumption that the target atom is isolated. While this assumption is acceptable for noble elements, as for argon or xenon detectors~\cite{Bell:2021ihi}, it is expected to be less valid in semiconductors, where solid-state effects should be considered. However, developing a first-principle theory that goes beyond the isolated atom approximation is challenging because of the many-body effects that need to be taken into account.
Remarkably, the formalism developed in Ref.~\cite{Liu:2020pat} relates the photoabsorption cross section $\sigma_\gamma$ to the dipole matrix element, necessary to compute the Migdal ionization rate, without requiring any many-body calculation.
This scheme will be referred to as Migdal photoabsorption approximation (MPA).
One of the major advantages of MPA is that the photoabsorption cross section is experimentally known, such that the Migdal rate suffers from very small uncertainties~\cite{Liu:2020pat}, well below the precision required in this work. 
MPA has been so far adopted in the context of dark matter searches, where the power of the formalism has been proved by comparing it to other computations for silicon and xenon~\cite{Ibe:2017yqa,Essig:2019xkx}. 
However, MPA has never been exploited in the context of neutrino scattering. Here, for the first time, we use it in this context and we compare its predictions with the formalism of Ibe \textit{et\,al.} for germanium detectors.
Explicitly, we derive the Migdal contribution to the \cenns cross section under MPA as (see Appendix~\ref{app:MigdalRate} for further information)
\begin{align}\nonumber
\left(\dfrac{d^2\sigma_{\bar{\nu}_e\text{-}\mathcal{N}}}{d T_\mathrm{nr}dE_r}\right)_{\rm{Migdal}}^{\rm{MPA}}
	 =& 
	\dfrac{G_{\text{F}}^2 M}{\pi}
	\left( 1 - \dfrac{M T_\mathrm{nr}}{2 E_\nu^2} \right)\mathcal{Q}_W^2\times\\
 &\frac{1}{2\pi^2\alpha_{\rm{EM}}}\frac{m^2_e}{M}\frac{T_{\rm{nr}}}{E_r}\sigma_\gamma^{\rm{Ge}}(E_r),
\label{eq:MigdalMPA}
\end{align}
where $\alpha_{\rm{EM}}$ is the fine structure constant and $E_r$ is the energy deposit due to atomic excitation or ionization such that $E_{\rm{det}}=f_{Q}T_{\rm{nr}}+E_r$.
The photoabsorption cross section $\sigma_\gamma^{\rm{Ge}}(E_r)$ for Ge has been taken from Refs.~\cite{HENKE1993181,Chen:2014ypv} for $E_r\geq10\;\textrm{eV}_\textrm{ee}$.
The theoretical event rate as a function of $E_{\rm{det}}$ is obtained by integrating over all possible nuclear recoil and neutrino energies and imposing energy conservation.

It should be noticed that depending on the crystal scale that one is able to probe, other effects that account for the response of multiple atoms at once should be considered, as they have been proved to highly enhance the Migdal rate~\cite{Knapen:2020aky,Knapen:2021bwg,Adams:2022zvg,Liang:2022xbu}. 
However, although the range of $T_{\rm{nr}}$ in Eq.~(\ref{eq:range}) includes very small values, the main contribution in current Ge \cenns reactor experiments comes from $T_{\rm{nr}}$ of the order of 1 $\mathrm{keV}$.
Thus, the momentum transfer is $|\vec{q}|\simeq\sqrt{2MT_{\rm{nr}}}\sim10\;\rm{MeV}$ with a corresponding de Broglie wavelength of about 20 fm. The latter is much smaller than the scale of the interparticle spacing in the crystal so, in this work, we can safely neglect these effects. We point here, that at lower ionization energies, multiple atom effects must be better taken into consideration. Moreover, it has been suggested that in germanium a larger amount of secondary nuclear recoils may be produced following a low-energy primary one~\cite{Collar:2021fcl}. This, in addition to possible more complex crystal response models ~\cite{Kahn:2020fef, Kurinsky:2020dpb, Trickle:2019nya}, can significantly affect the microscopic description of the Migdal effect. However, the inclusion of these possible effects, which still need to be investigated both theoretically and experimentally, is beyond the scope of this work.

\section{Results and Discussion}

In order to compare the impact of the Migdal contribution to the standard \cenns rate,  we consider a 1 kg Ge detector located $10\,\rm{m}$ away from a $3\,\rm{GW}_{\rm{th}}$ reactor power plant whose $\bar{\nu}_e$ spectrum is given by the HMVE parameterization. This specific configuration resembles that of current \cenns reactor experiments, like Dresden-II, CONUS and $\nu$GEN. 

\begin{figure}[tb]
    \centering
    \includegraphics[width=\columnwidth]{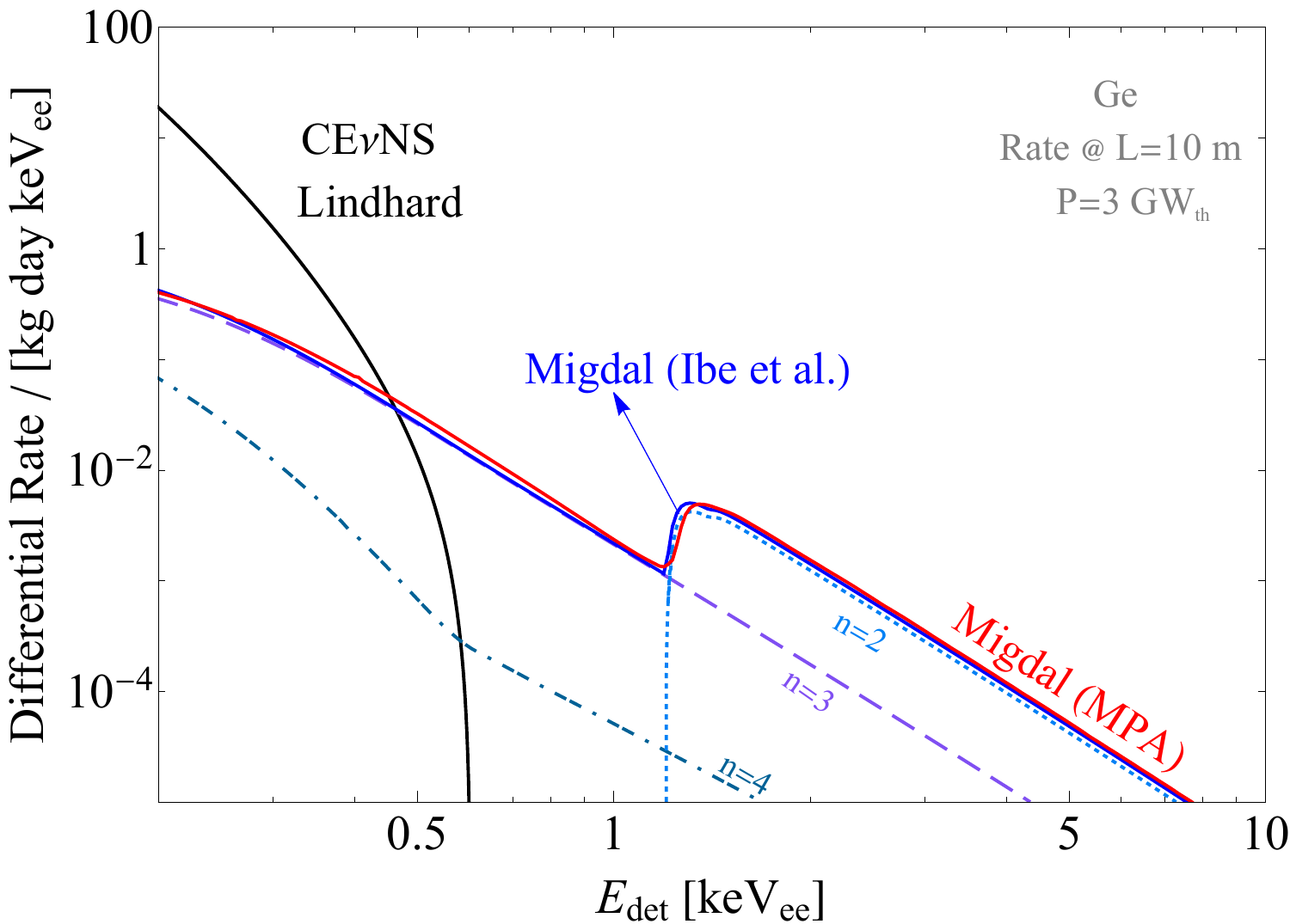}
    \caption{\cenns theoretical differential rate for a Ge detector located $10\,\rm{m}$ away from a $3\,\rm{GW}_{\rm{th}}$ reactor using the Lindhard QF (black solid line). We also show the Migdal rate obtained under the Ibe \textit{et al.} formalism (solid blue line), highlighting also the contributions of the different $n=2,3,4$ atomic shells (dashed curves), and the rate obtained with the MPA formalism (solid red line).}
    \label{fig:PlotMigdal}
\end{figure}

Under these assumptions, in Fig.~\ref{fig:PlotMigdal} we show the theoretical \cenns rate obtained with Eq.~(\ref{eq:ratecevns}) as a function of the detected energy, adopting for the QF the standard Lindhard parameterization.
We also show the Migdal rate determined with the Ibe \textit{et al.} formalism as in Eq.~(\ref{eq:rateMigdal}), isolating the contributions from the different $n$ shells, obtained by summing over all the contributions from different orbital angular momenta in the initial state $\ell$. We compare this result with the Migdal prediction under the MPA scheme, finding intriguingly that the two formalisms give practically identical results in the energy range considered. Moreover, in both cases, the Migdal contribution is completely subdominant with respect to the \cenns one for energies below $\sim0.6\;\textrm{keV}_{\rm{ee}}$, with the most significant contribution given by the $n=2,\,3$ shells. Above $\sim0.6\;\textrm{keV}_{\rm{ee}}$ it starts to dominate\footnote{
Note that a similar trend is also found in Ref.~\cite{Bell:2021ihi}, where a comparison between the \cenns rate and the Migdal contribution using the formalism of Ibe \textit{et\,al.} has been evaluated for xenon and argon detectors in a reactor site. We have used their publicly available code~\cite{Bell:2021ihi} to verify our results.}, and it could provide the possibility to observe \cenns above this threshold, even if being so small it would require extremely low levels of background.

\begin{figure}[tb]
    \includegraphics[width=\columnwidth]{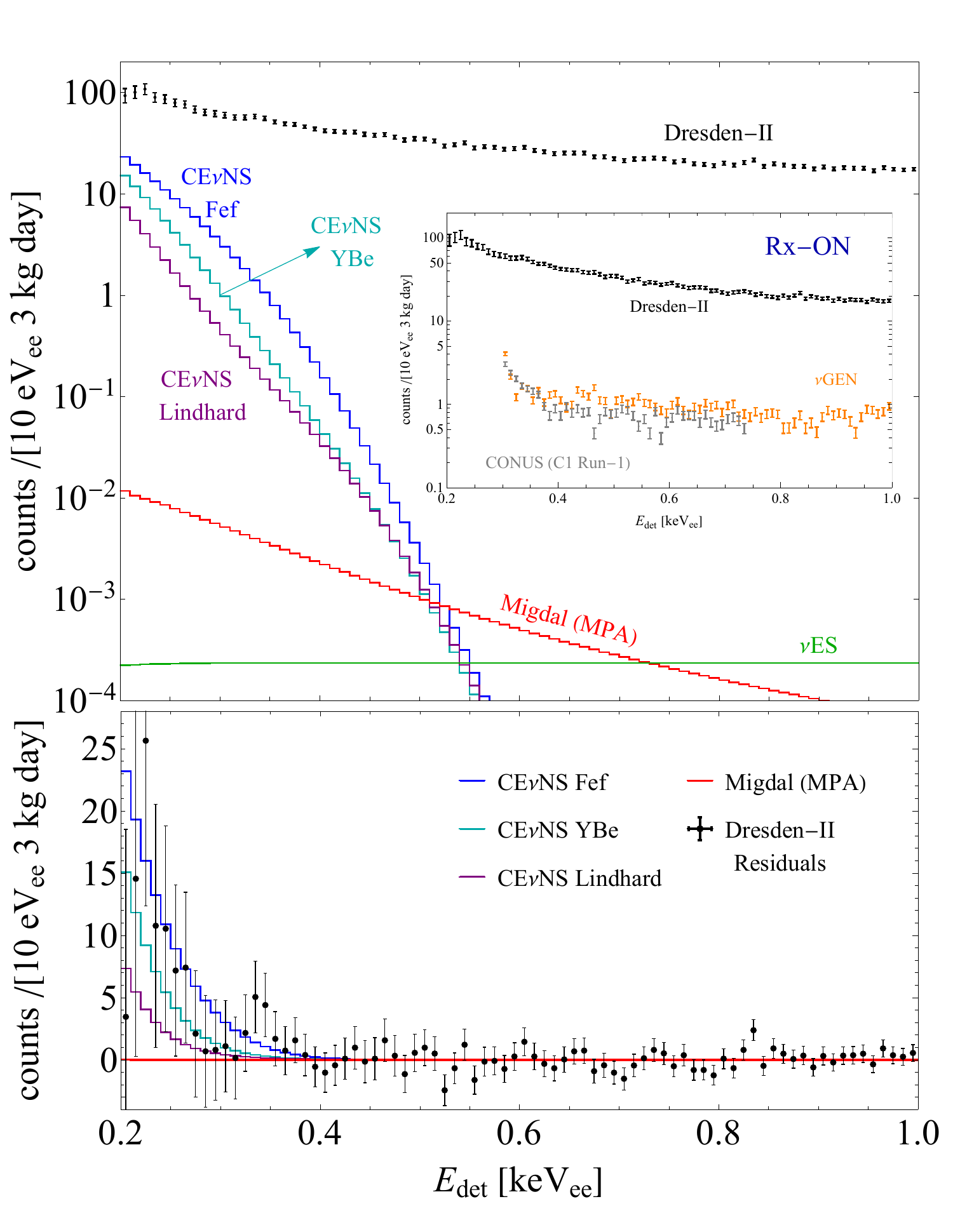}
    \caption{Expected number of \cenns events in the Dresden-II detector obtained using different quenching factors, i.e. Fef (blue line), YBe (cyan line) and Lindhard (purple line). The Migdal contribution corresponds to the red curve, while the neutrino electron scattering ($\nu$ES) contribution is given by the green line. In the top panel, we compare these curves with the Dresden-II reactor ON data, while in the bottom panel, we show the Dresden-II data residuals after background subtraction. The inset shows a comparison of Dresden-II reactor ON (Rx-ON)~\cite{Colaresi:2022obx}, $\nu$GEN~\cite{nGeN:2022uje} and CONUS (C1 Run-1)~\cite{CONUS:2020skt} data, all rescaled to the same units.}
    \label{fig:MigDII}
\end{figure}

We now focus on the Dresden-II science case. Here, as already stated in the introduction, the Migdal effect has been invoked as a possible explanation of the enhancement measured in the Fef and YBe quenching factors at low energies that in turn enabled the observation of \cenns in the Dresden-II data. 
In the top panel of Fig.~\ref{fig:MigDII}, we show the Dresden-II reactor ON (Rx-ON) data along with the standard \cenns predictions obtained with three different QFs, namely Lindhard, Fef and YBe. To derive these spectra we used all the experimental information on the Dresden-II detector, including energy-smearing effects, following Refs.~\cite{Colaresi:2022obx,Coloma:2022avw,AtzoriCorona:2022qrf}. In the bottom panel of Fig.~\ref{fig:MigDII} we show the same spectra for the three QFs but we compare them to the Dresden-II data residuals after background subtraction~\cite{Colaresi:2022obx}. It is evident that only the Fef and marginally the YBe QFs fit the excess and lead to a statistically significant \cenns observation for $E_{\rm{det}}\lesssim0.3\,\mathrm{keV}_\mathrm{ee}$. In the same figure, we also show the Migdal contribution using the MPA formalism.
It is clear that adding the Migdal contribution to the \cenns Lindhard prediction is not sufficient to explain the \cenns Fef or YBe predictions, given that the former is completely negligible with respect to the \cenns signal. Moreover, we find that the neutrino-electron scattering ($\nu$ES)~\cite{AtzoriCorona:2022qrf,AtzoriCorona:2022jeb} overcomes the Migdal rate for $E_{\rm{det}}\gtrsim0.7\;\rm{keV}_{\rm{ee}}$, as shown in the top panel of Fig.~\ref{fig:MigDII}. Overall, both Migdal and $\nu$ES rates are so small that with the current experimental precision can be overlooked in SM \cenns searches. However, in some scenarios of physics beyond the SM, like non-standard properties of neutrinos, their contribution could be significantly enhanced and thus they need to be taken into account~\cite{Coloma:2022avw, AtzoriCorona:2022qrf} to derive meaningful limits.

In the inset of Fig.~\ref{fig:MigDII}, for comparison purposes, we show a review of the existing data from germanium detectors searching for \cenns at a reactor site, i.e. Dresden-II~\cite{Colaresi:2022obx}, $\nu$GEN~\cite{nGeN:2022uje} and CONUS (C1 Run-1)~\cite{CONUS:2020skt}.  
Interestingly, despite the fact that CONUS and $\nu$Gen have reached a much lower background level compared to Dresden-II, they have not detected \cenns yet. Having a conservative threshold of $\sim0.3\;\rm{keV}_{\rm{ee}}$, future data will be needed to confirm the  Dresden-II observation of \cenns, which is localized below this threshold. Nevertheless, despite the low background level reached, the Migdal contribution is so small that it could be safely neglected also in experiments like $\nu$GEN and CONUS, which show a good agreement with the expected background.  Similar conclusions are expected also for silicon detectors like CONNIE~\cite{CONNIE:2016nav,CONNIE:2021ggh} that operate in a similar energy range.

To conclude, we have shown that the Migdal contribution is orders-of-magnitude subdominant in the region of interest for reactor \cenns searches with germanium detectors, independently of the formalism used to model the Migdal effect.
Thus, the enhancement of the quenching factor at low energies found in Ref.~\cite{Colaresi:2022obx} that enabled the observation of \cenns at the Dresden-II site requires a different explanation than the standard Migdal effect.

\begin{acknowledgements}
The authors are thankful to Simon Knapen for a fruitful discussion on the topic. The work of C. Giunti is supported by the PRIN 2022 research grant “Addressing systematic uncertainties in searches for dark matter”, Number 2022F2843L, funded by MIUR.

\end{acknowledgements}

\appendix
\section{Derivation of the Migdal cross section with the MPA formalism}
\label{app:MigdalRate}

While the standard Migdal formalism derived by Ibe \textit{et al.} has been already discussed in previous studies in the context of \cenns interactions (see e.g. Ref.~\cite{Ibe:2017yqa} and Refs.~\cite{Bell:2019egg,Bell:2021ihi} with their publicly-available codes), little literature is present for the so-called Migdal photoabsorption approximation (MPA) as this has never been used in this context. In this Appendix, we expand the discussion on the MPA formalism providing the reader with more details. In this scenario, the Migdal transition matrix element under the dipole approximation, $|M_{fi}|^2$, is related to the ionization probability due to the Migdal effect. It is convenient to define the double differential cross section for an electron neutrino that scatters off a nucleus $\mathcal{N}$ as
\begin{align}
\left(\dfrac{d^2\sigma_{\bar{\nu}_e\text{-}\mathcal{N}}}{d T_\mathrm{nr}dE_r}\right)_{\rm{Migdal}}
	 =& 
	\left(\dfrac{d\sigma_{\bar{\nu}_e\text{-}\mathcal{N}}}{d T_\mathrm{nr}}\right)|M_{fi}|^2,
\label{eq:CrossSecdifferential}
\end{align}
where $T_{\rm{nr}}$ is the nuclear recoil energy, and the matrix element is a function of the energy deposit $E_r$.
The square of the transition matrix element in Eq.~(\ref{eq:CrossSecdifferential}) can be recast to
\begin{align}
    |M_{fi}|^2=
    \left(\frac{m_e}{M}\right)^2 2M T_{\rm{nr}}\overline{D^2_{fi}},
    \label{eq:Mfisquare}
\end{align}
where $M$ is the nuclear mass, $m_e$ is the electron mass and we have used the fact that $M |\vec{v}|=\sqrt{ 2 M T_{\rm{nr}}}$ is the momentum transfer of the recoiling nucleus, with $\vec{v}$ the nuclear recoil velocity. Here, $\overline{D^2_{fi}}$ is the average of the squared dipole matrix element.
Under the MPA scheme, one has to relate $\overline{D^2_{fi}}$ with the photoabsorption cross section $\sigma_\gamma$ through~\cite{Liu:2020pat}
\begin{align}
    \overline{D^2_{fi}}=
    \frac{\sigma_\gamma(E_r)}{4\pi^2\alpha_{\rm{EM}} E_r},
    \label{eq:MPApproximation}
\end{align}
where $\alpha_{\rm{EM}}$ is the fine structure constant.
\begin{figure}[!t]
    \includegraphics[width=0.9\columnwidth]{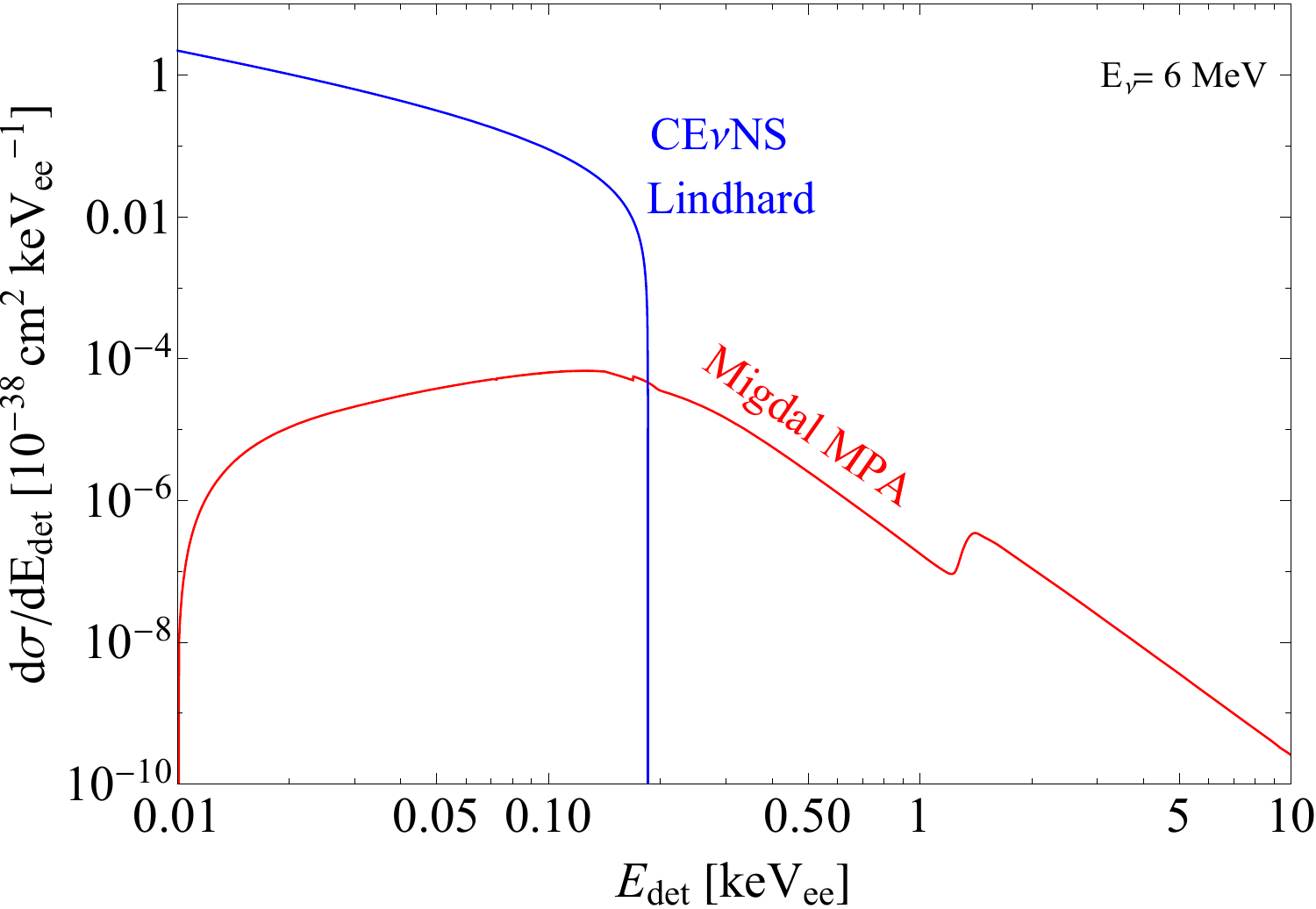}
    \caption{Comparison between the Migdal cross section (under the MPA scheme) and the \cenns one as a function of the observed energy $E_{\rm{det}}$, assuming the standard Lindhard quenching factor and a neutrino energy of $E_\nu= 6 $ MeV.}
    \label{fig:MigCrossSec}
\end{figure}

In Fig.~\ref{fig:MigCrossSec} we show the comparison between the differential Migdal cross section and the \cenns one as a function of the observed energy $E_{\rm{det}}$, for a neutrino energy of $E_\nu=6\; \rm{MeV}$ and assuming the standard Lindhard quenching factor.
By substituting Eq.~(\ref{eq:MPApproximation}) into Eq.~(\ref{eq:Mfisquare}) and using Eq.~(\ref{eq:CrossSecdifferential}) one obtains Eq.~(11). 

\begin{figure}[t]
    \includegraphics[width=1.1\columnwidth]{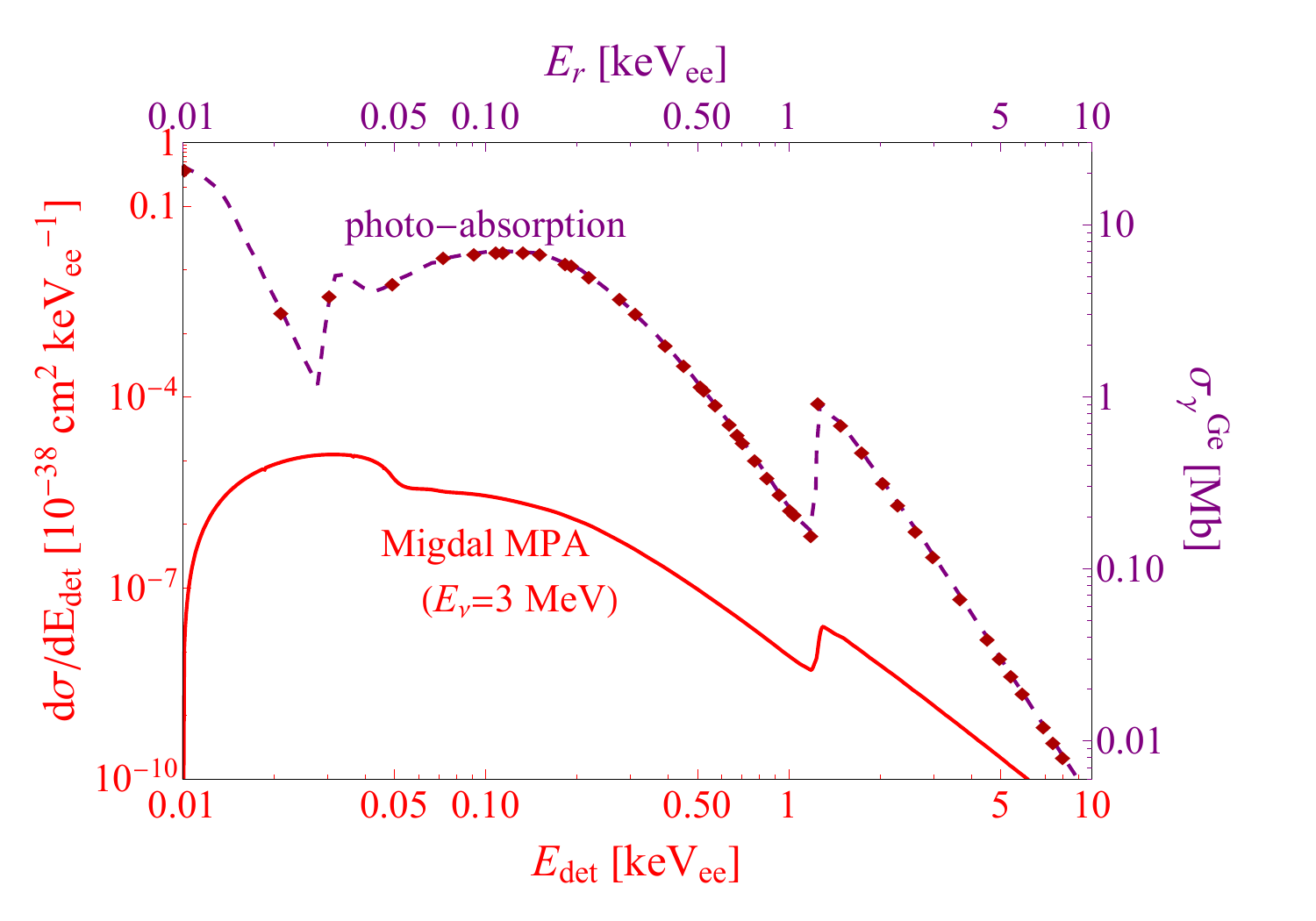}
    \caption{Photoabsorption cross section for Germanium taken from Ref.~\cite{HENKE1993181} (purple) and differential Migdal cross section for $E_\nu=3\;$MeV. The dots represent the photoabsorption cross section for each x-ray emission as listed in Ref.~\cite{HENKE1993181}.}
    \label{fig:sigmagammaGeDouble}
\end{figure}

In Fig.~\ref{fig:sigmagammaGeDouble} we show the germanium photoabsorption cross section extracted from Ref.~\cite{HENKE1993181} as used in this work, among with the differential Migdal cross section as a function of $E_{\rm{det}}$ for $E_\nu=3\;$MeV. 
The two cross sections are reported at different energy scales indicated in the corresponding $y$ axes.

The differential Migdal cross section as a function of $E_{\rm{det}}$ is obtained by integrating over all possible nuclear recoils and is given by
\begin{align}\nonumber
&\left(\dfrac{d\sigma_{\bar{\nu}_e\text{-}\mathcal{N}}}{dE_{\rm{det}}}\right)_{\rm{Migdal}}^{\rm{MPA}}
	 =
	\dfrac{G_{\text{F}}^2 M}{\pi}
	\int_{T_\textrm{nr}^{\rm{min}}}^{T_{\rm{nr}}^{\rm{max}}} dT_{\rm{nr}}\left( 1 - \dfrac{M T_\mathrm{nr}}{2 E_\nu^2} \right)\mathcal{Q}_W^2\times\\
 &\frac{1}{2\pi^2\alpha_{\rm{EM}}}\frac{m^2_e}{M}\frac{T_{\rm{nr}}}{E_{\rm{det}}-f_QT_{\rm{nr}}}\sigma_\gamma^{\rm{Ge}}(E_{\rm{det}}-f_QT_{\rm{nr}}),
\label{eq:MigdalMPA2}
\end{align}
where $G_{\text{F}}$ is the Fermi constant, $\mathcal{Q}_W$ is the weak charge of the nucleus and $f_Q$ is the energy-dependent parameterization of the quenching factor. 
In Fig.~\ref{fig:sigmagammaGeDouble}, it is possible to see that the features that characterize the differential Migdal cross section within the MPA scheme are tightly connected with the photoabsorption peaks.

\bibliography{ref}

\begin{thebibliography}{84}%
\makeatletter
\providecommand \@ifxundefined [1]{%
 \@ifx{#1\undefined}
}%
\providecommand \@ifnum [1]{%
 \ifnum #1\expandafter \@firstoftwo
 \else \expandafter \@secondoftwo
 \fi
}%
\providecommand \@ifx [1]{%
 \ifx #1\expandafter \@firstoftwo
 \else \expandafter \@secondoftwo
 \fi
}%
\providecommand \natexlab [1]{#1}%
\providecommand \enquote  [1]{``#1''}%
\providecommand \bibnamefont  [1]{#1}%
\providecommand \bibfnamefont [1]{#1}%
\providecommand \citenamefont [1]{#1}%
\providecommand \href@noop [0]{\@secondoftwo}%
\providecommand \href [0]{\begingroup \@sanitize@url \@href}%
\providecommand \@href[1]{\@@startlink{#1}\@@href}%
\providecommand \@@href[1]{\endgroup#1\@@endlink}%
\providecommand \@sanitize@url [0]{\catcode `\\12\catcode `\$12\catcode `\&12\catcode `\#12\catcode `\^12\catcode `\_12\catcode `\%12\relax}%
\providecommand \@@startlink[1]{}%
\providecommand \@@endlink[0]{}%
\providecommand \url  [0]{\begingroup\@sanitize@url \@url }%
\providecommand \@url [1]{\endgroup\@href {#1}{\urlprefix }}%
\providecommand \urlprefix  [0]{URL }%
\providecommand \Eprint [0]{\href }%
\providecommand \doibase [0]{https://doi.org/}%
\providecommand \selectlanguage [0]{\@gobble}%
\providecommand \bibinfo  [0]{\@secondoftwo}%
\providecommand \bibfield  [0]{\@secondoftwo}%
\providecommand \translation [1]{[#1]}%
\providecommand \BibitemOpen [0]{}%
\providecommand \bibitemStop [0]{}%
\providecommand \bibitemNoStop [0]{.\EOS\space}%
\providecommand \EOS [0]{\spacefactor3000\relax}%
\providecommand \BibitemShut  [1]{\csname bibitem#1\endcsname}%
\let\auto@bib@innerbib\@empty
\bibitem [{\citenamefont {Freedman}(1974)}]{Freedman:1973yd}%
  \BibitemOpen
  \bibfield  {author} {\bibinfo {author} {\bibfnamefont {D.~Z.}\ \bibnamefont {Freedman}},\ }\bibfield  {title} {\bibinfo {title} {Coherent effects of a weak neutral current},\ }\href {https://doi.org/10.1103/PhysRevD.9.1389} {\bibfield  {journal} {\bibinfo  {journal} {Phys. Rev. D}\ }\textbf {\bibinfo {volume} {9}},\ \bibinfo {pages} {1389} (\bibinfo {year} {1974})}\BibitemShut {NoStop}%
\bibitem [{\citenamefont {Akimov}\ \emph {et~al.}(2017)\citenamefont {Akimov} \emph {et~al.}}]{COHERENT:2017ipa}%
  \BibitemOpen
  \bibfield  {author} {\bibinfo {author} {\bibfnamefont {D.}~\bibnamefont {Akimov}} \emph {et~al.} (\bibinfo {collaboration} {COHERENT}),\ }\bibfield  {title} {\bibinfo {title} {{Observation of Coherent Elastic Neutrino-Nucleus Scattering}},\ }\href {https://doi.org/10.1126/science.aao0990} {\bibfield  {journal} {\bibinfo  {journal} {Science}\ }\textbf {\bibinfo {volume} {357}},\ \bibinfo {pages} {1123} (\bibinfo {year} {2017})},\ \Eprint {https://arxiv.org/abs/1708.01294} {arXiv:1708.01294 [nucl-ex]} \BibitemShut {NoStop}%
\bibitem [{\citenamefont {Akimov}\ \emph {et~al.}(2021)\citenamefont {Akimov} \emph {et~al.}}]{COHERENT:2020iec}%
  \BibitemOpen
  \bibfield  {author} {\bibinfo {author} {\bibfnamefont {D.}~\bibnamefont {Akimov}} \emph {et~al.} (\bibinfo {collaboration} {COHERENT}),\ }\bibfield  {title} {\bibinfo {title} {{First Detection of Coherent Elastic Neutrino-Nucleus Scattering on Argon}},\ }\href@noop {} {\bibfield  {journal} {\bibinfo  {journal} {Phys.Rev.Lett.}\ }\textbf {\bibinfo {volume} {126}},\ \bibinfo {pages} {012002} (\bibinfo {year} {2021})},\ \Eprint {https://arxiv.org/abs/arXiv:2003.10630} {arXiv:2003.10630 [nucl-ex]} \BibitemShut {NoStop}%
\bibitem [{\citenamefont {Akimov}\ \emph {et~al.}(2022{\natexlab{a}})\citenamefont {Akimov} \emph {et~al.}}]{COHERENT:2021xmm}%
  \BibitemOpen
  \bibfield  {author} {\bibinfo {author} {\bibfnamefont {D.}~\bibnamefont {Akimov}} \emph {et~al.} (\bibinfo {collaboration} {COHERENT}),\ }\bibfield  {title} {\bibinfo {title} {{Measurement of the Coherent Elastic Neutrino-Nucleus Scattering Cross Section on CsI by COHERENT}},\ }\href@noop {} {\bibfield  {journal} {\bibinfo  {journal} {Phys.Rev.Lett.}\ }\textbf {\bibinfo {volume} {129}},\ \bibinfo {pages} {081801} (\bibinfo {year} {2022}{\natexlab{a}})},\ \Eprint {https://arxiv.org/abs/arXiv:2110.07730} {arXiv:2110.07730 [hep-ex]} \BibitemShut {NoStop}%
\bibitem [{\citenamefont {Angloher}\ \emph {et~al.}(2019)\citenamefont {Angloher} \emph {et~al.}}]{NUCLEUS:2019igx}%
  \BibitemOpen
  \bibfield  {author} {\bibinfo {author} {\bibfnamefont {G.}~\bibnamefont {Angloher}} \emph {et~al.} (\bibinfo {collaboration} {NUCLEUS}),\ }\bibfield  {title} {\bibinfo {title} {{Exploring $\hbox {CE}\nu \hbox {NS}$ with NUCLEUS at the Chooz nuclear power plant}},\ }\href {https://doi.org/10.1140/epjc/s10052-019-7454-4} {\bibfield  {journal} {\bibinfo  {journal} {Eur. Phys. J. C}\ }\textbf {\bibinfo {volume} {79}},\ \bibinfo {pages} {1018} (\bibinfo {year} {2019})},\ \Eprint {https://arxiv.org/abs/1905.10258} {arXiv:1905.10258 [physics.ins-det]} \BibitemShut {NoStop}%
\bibitem [{\citenamefont {Strauss}\ \emph {et~al.}(2017)\citenamefont {Strauss} \emph {et~al.}}]{Strauss:2017cuu}%
  \BibitemOpen
  \bibfield  {author} {\bibinfo {author} {\bibfnamefont {R.}~\bibnamefont {Strauss}} \emph {et~al.},\ }\bibfield  {title} {\bibinfo {title} {{The $\nu$-cleus experiment: A gram-scale fiducial-volume cryogenic detector for the first detection of coherent neutrino-nucleus scattering}},\ }\href {https://doi.org/10.1140/epjc/s10052-017-5068-2} {\bibfield  {journal} {\bibinfo  {journal} {Eur. Phys. J. C}\ }\textbf {\bibinfo {volume} {77}},\ \bibinfo {pages} {506} (\bibinfo {year} {2017})},\ \Eprint {https://arxiv.org/abs/1704.04320} {arXiv:1704.04320 [physics.ins-det]} \BibitemShut {NoStop}%
\bibitem [{\citenamefont {Akimov}\ \emph {et~al.}(2022{\natexlab{b}})\citenamefont {Akimov} \emph {et~al.}}]{Akimov:2022oyb}%
  \BibitemOpen
  \bibfield  {author} {\bibinfo {author} {\bibfnamefont {D.}~\bibnamefont {Akimov}} \emph {et~al.},\ }\bibfield  {title} {\bibinfo {title} {{The COHERENT Experimental Program}},\ }in\ \href@noop {} {\emph {\bibinfo {booktitle} {{Snowmass 2021}}}}\ (\bibinfo {year} {2022})\ \Eprint {https://arxiv.org/abs/2204.04575} {arXiv:2204.04575 [hep-ex]} \BibitemShut {NoStop}%
\bibitem [{\citenamefont {Baxter}\ \emph {et~al.}(2020)\citenamefont {Baxter} \emph {et~al.}}]{Baxter:2019mcx}%
  \BibitemOpen
  \bibfield  {author} {\bibinfo {author} {\bibfnamefont {D.}~\bibnamefont {Baxter}} \emph {et~al.},\ }\bibfield  {title} {\bibinfo {title} {{Coherent Elastic Neutrino-Nucleus Scattering at the European Spallation Source}},\ }\href {https://doi.org/10.1007/JHEP02(2020)123} {\bibfield  {journal} {\bibinfo  {journal} {JHEP}\ }\textbf {\bibinfo {volume} {02}},\ \bibinfo {pages} {123}},\ \Eprint {https://arxiv.org/abs/1911.00762} {arXiv:1911.00762 [physics.ins-det]} \BibitemShut {NoStop}%
\bibitem [{\citenamefont {Bonet}\ \emph {et~al.}(2021{\natexlab{a}})\citenamefont {Bonet} \emph {et~al.}}]{CONUS:2020skt}%
  \BibitemOpen
  \bibfield  {author} {\bibinfo {author} {\bibfnamefont {H.}~\bibnamefont {Bonet}} \emph {et~al.} (\bibinfo {collaboration} {CONUS}),\ }\bibfield  {title} {\bibinfo {title} {{Constraints on elastic neutrino nucleus scattering in the fully coherent regime from the CONUS experiment}},\ }\href {https://doi.org/10.1103/PhysRevLett.126.041804} {\bibfield  {journal} {\bibinfo  {journal} {Phys. Rev. Lett.}\ }\textbf {\bibinfo {volume} {126}},\ \bibinfo {pages} {041804} (\bibinfo {year} {2021}{\natexlab{a}})},\ \Eprint {https://arxiv.org/abs/2011.00210} {arXiv:2011.00210 [hep-ex]} \BibitemShut {NoStop}%
\bibitem [{\citenamefont {Aguilar-Arevalo}\ \emph {et~al.}(2016)\citenamefont {Aguilar-Arevalo} \emph {et~al.}}]{CONNIE:2016nav}%
  \BibitemOpen
  \bibfield  {author} {\bibinfo {author} {\bibfnamefont {A.}~\bibnamefont {Aguilar-Arevalo}} \emph {et~al.} (\bibinfo {collaboration} {CONNIE}),\ }\bibfield  {title} {\bibinfo {title} {{Results of the Engineering Run of the Coherent Neutrino Nucleus Interaction Experiment (CONNIE)}},\ }\href {https://doi.org/10.1088/1748-0221/11/07/P07024} {\bibfield  {journal} {\bibinfo  {journal} {JINST}\ }\textbf {\bibinfo {volume} {11}}\bibfield  {number} {\bibinfo  {number} { (07)},\ \bibinfo {pages} {P07024}},\ }\Eprint {https://arxiv.org/abs/1604.01343} {arXiv:1604.01343 [physics.ins-det]} \BibitemShut {NoStop}%
\bibitem [{\citenamefont {Billard}\ \emph {et~al.}(2017)\citenamefont {Billard} \emph {et~al.}}]{Billard:2016giu}%
  \BibitemOpen
  \bibfield  {author} {\bibinfo {author} {\bibfnamefont {J.}~\bibnamefont {Billard}} \emph {et~al.},\ }\bibfield  {title} {\bibinfo {title} {{Coherent Neutrino Scattering with Low Temperature Bolometers at Chooz Reactor Complex}},\ }\href {https://doi.org/10.1088/1361-6471/aa83d0} {\bibfield  {journal} {\bibinfo  {journal} {J. Phys. G}\ }\textbf {\bibinfo {volume} {44}},\ \bibinfo {pages} {105101} (\bibinfo {year} {2017})},\ \Eprint {https://arxiv.org/abs/1612.09035} {arXiv:1612.09035 [physics.ins-det]} \BibitemShut {NoStop}%
\bibitem [{\citenamefont {Alekseev}\ \emph {et~al.}(2022)\citenamefont {Alekseev} \emph {et~al.}}]{nGeN:2022uje}%
  \BibitemOpen
  \bibfield  {author} {\bibinfo {author} {\bibfnamefont {I.}~\bibnamefont {Alekseev}} \emph {et~al.} (\bibinfo {collaboration} {\ensuremath{\nu}GeN}),\ }\bibfield  {title} {\bibinfo {title} {{First results of the \ensuremath{\nu}GeN experiment on coherent elastic neutrino-nucleus scattering}},\ }\href {https://doi.org/10.1103/PhysRevD.106.L051101} {\bibfield  {journal} {\bibinfo  {journal} {Phys. Rev. D}\ }\textbf {\bibinfo {volume} {106}},\ \bibinfo {pages} {L051101} (\bibinfo {year} {2022})},\ \Eprint {https://arxiv.org/abs/2205.04305} {arXiv:2205.04305 [nucl-ex]} \BibitemShut {NoStop}%
\bibitem [{\citenamefont {Colaresi}\ \emph {et~al.}(2021)\citenamefont {Colaresi}, \citenamefont {Collar}, \citenamefont {Hossbach}, \citenamefont {Kavner}, \citenamefont {Lewis}, \citenamefont {Robinson},\ and\ \citenamefont {Yocum}}]{Colaresi:2021kus}%
  \BibitemOpen
  \bibfield  {author} {\bibinfo {author} {\bibfnamefont {J.}~\bibnamefont {Colaresi}}, \bibinfo {author} {\bibfnamefont {J.~I.}\ \bibnamefont {Collar}}, \bibinfo {author} {\bibfnamefont {T.~W.}\ \bibnamefont {Hossbach}}, \bibinfo {author} {\bibfnamefont {A.~R.~L.}\ \bibnamefont {Kavner}}, \bibinfo {author} {\bibfnamefont {C.~M.}\ \bibnamefont {Lewis}}, \bibinfo {author} {\bibfnamefont {A.~E.}\ \bibnamefont {Robinson}},\ and\ \bibinfo {author} {\bibfnamefont {K.~M.}\ \bibnamefont {Yocum}},\ }\bibfield  {title} {\bibinfo {title} {{First results from a search for coherent elastic neutrino-nucleus scattering at a reactor site}},\ }\href {https://doi.org/10.1103/PhysRevD.104.072003} {\bibfield  {journal} {\bibinfo  {journal} {Phys. Rev. D}\ }\textbf {\bibinfo {volume} {104}},\ \bibinfo {pages} {072003} (\bibinfo {year} {2021})},\ \Eprint {https://arxiv.org/abs/2108.02880} {arXiv:2108.02880 [hep-ex]} \BibitemShut {NoStop}%
\bibitem [{\citenamefont {Colaresi}\ \emph {et~al.}(2022)\citenamefont {Colaresi}, \citenamefont {Collar}, \citenamefont {Hossbach}, \citenamefont {Lewis},\ and\ \citenamefont {Yocum}}]{Colaresi:2022obx}%
  \BibitemOpen
  \bibfield  {author} {\bibinfo {author} {\bibfnamefont {J.}~\bibnamefont {Colaresi}}, \bibinfo {author} {\bibfnamefont {J.~I.}\ \bibnamefont {Collar}}, \bibinfo {author} {\bibfnamefont {T.~W.}\ \bibnamefont {Hossbach}}, \bibinfo {author} {\bibfnamefont {C.~M.}\ \bibnamefont {Lewis}},\ and\ \bibinfo {author} {\bibfnamefont {K.~M.}\ \bibnamefont {Yocum}},\ }\bibfield  {title} {\bibinfo {title} {{Measurement of Coherent Elastic Neutrino-Nucleus Scattering from Reactor Antineutrinos}},\ }\href {https://doi.org/10.1103/PhysRevLett.129.211802} {\bibfield  {journal} {\bibinfo  {journal} {Phys. Rev. Lett.}\ }\textbf {\bibinfo {volume} {129}},\ \bibinfo {pages} {211802} (\bibinfo {year} {2022})},\ \Eprint {https://arxiv.org/abs/2202.09672} {arXiv:2202.09672 [hep-ex]} \BibitemShut {NoStop}%
\bibitem [{\citenamefont {Aguilar-Arevalo}\ \emph {et~al.}(2022)\citenamefont {Aguilar-Arevalo} \emph {et~al.}}]{CONNIE:2021ggh}%
  \BibitemOpen
  \bibfield  {author} {\bibinfo {author} {\bibfnamefont {A.}~\bibnamefont {Aguilar-Arevalo}} \emph {et~al.} (\bibinfo {collaboration} {CONNIE}),\ }\bibfield  {title} {\bibinfo {title} {{Search for coherent elastic neutrino-nucleus scattering at a nuclear reactor with CONNIE 2019 data}},\ }\href {https://doi.org/10.1007/JHEP05(2022)017} {\bibfield  {journal} {\bibinfo  {journal} {JHEP}\ }\textbf {\bibinfo {volume} {05}},\ \bibinfo {pages} {017}},\ \Eprint {https://arxiv.org/abs/2110.13033} {arXiv:2110.13033 [hep-ex]} \BibitemShut {NoStop}%
\bibitem [{\citenamefont {Agnolet}\ \emph {et~al.}(2017)\citenamefont {Agnolet} \emph {et~al.}}]{MINER:2016igy}%
  \BibitemOpen
  \bibfield  {author} {\bibinfo {author} {\bibfnamefont {G.}~\bibnamefont {Agnolet}} \emph {et~al.} (\bibinfo {collaboration} {MINER}),\ }\bibfield  {title} {\bibinfo {title} {{Background Studies for the MINER Coherent Neutrino Scattering Reactor Experiment}},\ }\href {https://doi.org/10.1016/j.nima.2017.02.024} {\bibfield  {journal} {\bibinfo  {journal} {Nucl. Instrum. Meth. A}\ }\textbf {\bibinfo {volume} {853}},\ \bibinfo {pages} {53} (\bibinfo {year} {2017})},\ \Eprint {https://arxiv.org/abs/1609.02066} {arXiv:1609.02066 [physics.ins-det]} \BibitemShut {NoStop}%
\bibitem [{\citenamefont {Su}\ \emph {et~al.}(2023)\citenamefont {Su}, \citenamefont {Liu},\ and\ \citenamefont {Liang}}]{Su:2023klh}%
  \BibitemOpen
  \bibfield  {author} {\bibinfo {author} {\bibfnamefont {C.}~\bibnamefont {Su}}, \bibinfo {author} {\bibfnamefont {Q.}~\bibnamefont {Liu}},\ and\ \bibinfo {author} {\bibfnamefont {T.}~\bibnamefont {Liang}} (\bibinfo {collaboration} {CE\ensuremath{\nu}NS@CSNS}),\ }\href@noop {} {\bibinfo {title} {{CE$\nu$NS Experiment Proposal at CSNS}}} (\bibinfo {year} {2023}),\ \Eprint {https://arxiv.org/abs/2303.13423} {arXiv:2303.13423 [physics.ins-det]} \BibitemShut {NoStop}%
\bibitem [{\citenamefont {Colas}\ \emph {et~al.}(2022)\citenamefont {Colas}, \citenamefont {Billard}, \citenamefont {Ferriol}, \citenamefont {Gascon},\ and\ \citenamefont {Salagnac}}]{Colas:2021pxr}%
  \BibitemOpen
  \bibfield  {author} {\bibinfo {author} {\bibfnamefont {J.}~\bibnamefont {Colas}}, \bibinfo {author} {\bibfnamefont {J.}~\bibnamefont {Billard}}, \bibinfo {author} {\bibfnamefont {S.}~\bibnamefont {Ferriol}}, \bibinfo {author} {\bibfnamefont {J.}~\bibnamefont {Gascon}},\ and\ \bibinfo {author} {\bibfnamefont {T.}~\bibnamefont {Salagnac}} (\bibinfo {collaboration} {RICOCHET}),\ }\bibfield  {title} {\bibinfo {title} {{Development of data processing and analysis pipeline for the RICOCHET experiment}},\ }\href@noop {} {\bibfield  {journal} {\bibinfo  {journal} {J.Low Temp.Phys.}\ } (\bibinfo {year} {2022})},\ \Eprint {https://arxiv.org/abs/arXiv:2111.12856} {arXiv:2111.12856 [physics.ins-det]} \BibitemShut {NoStop}%
\bibitem [{\citenamefont {Akimov}\ \emph {et~al.}(2022{\natexlab{c}})\citenamefont {Akimov} \emph {et~al.}}]{Akimov:2022xvr}%
  \BibitemOpen
  \bibfield  {author} {\bibinfo {author} {\bibfnamefont {D.~Y.}\ \bibnamefont {Akimov}} \emph {et~al.},\ }\bibfield  {title} {\bibinfo {title} {{The RED-100 experiment}},\ }\href {https://doi.org/10.1088/1748-0221/17/11/T11011} {\bibfield  {journal} {\bibinfo  {journal} {JINST}\ }\textbf {\bibinfo {volume} {17}}\bibfield  {number} {\bibinfo  {number} { (11)},\ \bibinfo {pages} {T11011}},\ }\Eprint {https://arxiv.org/abs/2209.15516} {arXiv:2209.15516 [physics.ins-det]} \BibitemShut {NoStop}%
\bibitem [{\citenamefont {Wong}(2015)}]{Wong:2015kgl}%
  \BibitemOpen
  \bibfield  {author} {\bibinfo {author} {\bibfnamefont {H.~T.-K.}\ \bibnamefont {Wong}},\ }\bibfield  {title} {\bibinfo {title} {{Taiwan EXperiment On NeutrinO \textemdash{} History and Prospects}},\ }\href {https://doi.org/10.1142/S0217751X18300144} {\bibfield  {journal} {\bibinfo  {journal} {The Universe}\ }\textbf {\bibinfo {volume} {3}},\ \bibinfo {pages} {22} (\bibinfo {year} {2015})},\ \Eprint {https://arxiv.org/abs/1608.00306} {arXiv:1608.00306 [hep-ex]} \BibitemShut {NoStop}%
\bibitem [{\citenamefont {Cadeddu}\ \emph {et~al.}(2021{\natexlab{a}})\citenamefont {Cadeddu}, \citenamefont {Cargioli}, \citenamefont {Dordei}, \citenamefont {Giunti}, \citenamefont {Li}, \citenamefont {Picciau},\ and\ \citenamefont {Zhang}}]{Cadeddu:2020nbr}%
  \BibitemOpen
  \bibfield  {author} {\bibinfo {author} {\bibfnamefont {M.}~\bibnamefont {Cadeddu}}, \bibinfo {author} {\bibfnamefont {N.}~\bibnamefont {Cargioli}}, \bibinfo {author} {\bibfnamefont {F.}~\bibnamefont {Dordei}}, \bibinfo {author} {\bibfnamefont {C.}~\bibnamefont {Giunti}}, \bibinfo {author} {\bibfnamefont {Y.}~\bibnamefont {Li}}, \bibinfo {author} {\bibfnamefont {E.}~\bibnamefont {Picciau}},\ and\ \bibinfo {author} {\bibfnamefont {Y.}~\bibnamefont {Zhang}},\ }\bibfield  {title} {\bibinfo {title} {{Constraints on light vector mediators through coherent elastic neutrino nucleus scattering data from COHERENT}},\ }\href@noop {} {\bibfield  {journal} {\bibinfo  {journal} {JHEP}\ }\textbf {\bibinfo {volume} {2101}},\ \bibinfo {pages} {116}},\ \Eprint {https://arxiv.org/abs/arXiv:2008.05022} {arXiv:2008.05022 [hep-ph]} \BibitemShut {NoStop}%
\bibitem [{\citenamefont {Cadeddu}\ and\ \citenamefont {Dordei}(2019)}]{Cadeddu:2018izq}%
  \BibitemOpen
  \bibfield  {author} {\bibinfo {author} {\bibfnamefont {M.}~\bibnamefont {Cadeddu}}\ and\ \bibinfo {author} {\bibfnamefont {F.}~\bibnamefont {Dordei}},\ }\bibfield  {title} {\bibinfo {title} {{Reinterpreting the weak mixing angle from atomic parity violation in view of the Cs neutron rms radius measurement from COHERENT}},\ }\href@noop {} {\bibfield  {journal} {\bibinfo  {journal} {Phys.Rev.}\ }\textbf {\bibinfo {volume} {D99}},\ \bibinfo {pages} {033010} (\bibinfo {year} {2019})},\ \Eprint {https://arxiv.org/abs/arXiv:1808.10202} {arXiv:1808.10202 [hep-ph]} \BibitemShut {NoStop}%
\bibitem [{\citenamefont {De~Romeri}\ \emph {et~al.}(2023)\citenamefont {De~Romeri}, \citenamefont {Miranda}, \citenamefont {Papoulias}, \citenamefont {Sanchez~Garcia}, \citenamefont {Tortola},\ and\ \citenamefont {Valle}}]{DeRomeri:2022twg}%
  \BibitemOpen
  \bibfield  {author} {\bibinfo {author} {\bibfnamefont {V.}~\bibnamefont {De~Romeri}}, \bibinfo {author} {\bibfnamefont {O.~G.}\ \bibnamefont {Miranda}}, \bibinfo {author} {\bibfnamefont {D.~K.}\ \bibnamefont {Papoulias}}, \bibinfo {author} {\bibfnamefont {G.}~\bibnamefont {Sanchez~Garcia}}, \bibinfo {author} {\bibfnamefont {M.}~\bibnamefont {Tortola}},\ and\ \bibinfo {author} {\bibfnamefont {J.~W.~F.}\ \bibnamefont {Valle}},\ }\bibfield  {title} {\bibinfo {title} {{Physics implications of a combined analysis of COHERENT CsI and LAr data}},\ }\href@noop {} {\bibfield  {journal} {\bibinfo  {journal} {JHEP}\ }\textbf {\bibinfo {volume} {04}},\ \bibinfo {pages} {035}},\ \Eprint {https://arxiv.org/abs/arXiv:2211.11905} {arXiv:2211.11905 [hep-ph]} \BibitemShut {NoStop}%
\bibitem [{\citenamefont {Giunti}(2020)}]{Giunti:2019xpr}%
  \BibitemOpen
  \bibfield  {author} {\bibinfo {author} {\bibfnamefont {C.}~\bibnamefont {Giunti}},\ }\bibfield  {title} {\bibinfo {title} {{General COHERENT Constraints on Neutrino Non-Standard Interactions}},\ }\href@noop {} {\bibfield  {journal} {\bibinfo  {journal} {Phys.Rev.}\ }\textbf {\bibinfo {volume} {D101}},\ \bibinfo {pages} {035039} (\bibinfo {year} {2020})},\ \Eprint {https://arxiv.org/abs/arXiv:1909.00466} {arXiv:1909.00466 [hep-ph]} \BibitemShut {NoStop}%
\bibitem [{\citenamefont {Lindner}\ \emph {et~al.}(2017)\citenamefont {Lindner}, \citenamefont {Rodejohann},\ and\ \citenamefont {Xu}}]{Lindner:2016wff}%
  \BibitemOpen
  \bibfield  {author} {\bibinfo {author} {\bibfnamefont {M.}~\bibnamefont {Lindner}}, \bibinfo {author} {\bibfnamefont {W.}~\bibnamefont {Rodejohann}},\ and\ \bibinfo {author} {\bibfnamefont {X.-J.}\ \bibnamefont {Xu}},\ }\bibfield  {title} {\bibinfo {title} {{Coherent Neutrino-Nucleus Scattering and new Neutrino Interactions}},\ }\href@noop {} {\bibfield  {journal} {\bibinfo  {journal} {JHEP}\ }\textbf {\bibinfo {volume} {1703}},\ \bibinfo {pages} {097}},\ \Eprint {https://arxiv.org/abs/arXiv:1612.04150} {arXiv:1612.04150 [hep-ph]} \BibitemShut {NoStop}%
\bibitem [{\citenamefont {Billard}\ \emph {et~al.}(2018)\citenamefont {Billard}, \citenamefont {Johnston},\ and\ \citenamefont {Kavanagh}}]{Billard:2018jnl}%
  \BibitemOpen
  \bibfield  {author} {\bibinfo {author} {\bibfnamefont {J.}~\bibnamefont {Billard}}, \bibinfo {author} {\bibfnamefont {J.}~\bibnamefont {Johnston}},\ and\ \bibinfo {author} {\bibfnamefont {B.~J.}\ \bibnamefont {Kavanagh}},\ }\bibfield  {title} {\bibinfo {title} {{Prospects for exploring New Physics in Coherent Elastic Neutrino-Nucleus Scattering}},\ }\href@noop {} {\bibfield  {journal} {\bibinfo  {journal} {JCAP}\ }\textbf {\bibinfo {volume} {1811}},\ \bibinfo {pages} {016}},\ \Eprint {https://arxiv.org/abs/arXiv:1805.01798} {arXiv:1805.01798 [hep-ph]} \BibitemShut {NoStop}%
\bibitem [{\citenamefont {Cadeddu}\ \emph {et~al.}(2018{\natexlab{a}})\citenamefont {Cadeddu}, \citenamefont {Giunti}, \citenamefont {Li},\ and\ \citenamefont {Zhang}}]{Cadeddu:2017etk}%
  \BibitemOpen
  \bibfield  {author} {\bibinfo {author} {\bibfnamefont {M.}~\bibnamefont {Cadeddu}}, \bibinfo {author} {\bibfnamefont {C.}~\bibnamefont {Giunti}}, \bibinfo {author} {\bibfnamefont {Y.~F.}\ \bibnamefont {Li}},\ and\ \bibinfo {author} {\bibfnamefont {Y.~Y.}\ \bibnamefont {Zhang}},\ }\bibfield  {title} {\bibinfo {title} {{Average CsI neutron density distribution from COHERENT data}},\ }\href@noop {} {\bibfield  {journal} {\bibinfo  {journal} {Phys.Rev.Lett.}\ }\textbf {\bibinfo {volume} {120}},\ \bibinfo {pages} {072501} (\bibinfo {year} {2018}{\natexlab{a}})},\ \Eprint {https://arxiv.org/abs/arXiv:1710.02730} {arXiv:1710.02730 [hep-ph]} \BibitemShut {NoStop}%
\bibitem [{\citenamefont {Cadeddu}\ \emph {et~al.}(2021{\natexlab{b}})\citenamefont {Cadeddu}, \citenamefont {Cargioli}, \citenamefont {Dordei}, \citenamefont {Giunti}, \citenamefont {Li}, \citenamefont {Picciau}, \citenamefont {Ternes},\ and\ \citenamefont {Zhang}}]{Cadeddu:2021ijh}%
  \BibitemOpen
  \bibfield  {author} {\bibinfo {author} {\bibfnamefont {M.}~\bibnamefont {Cadeddu}}, \bibinfo {author} {\bibfnamefont {N.}~\bibnamefont {Cargioli}}, \bibinfo {author} {\bibfnamefont {F.}~\bibnamefont {Dordei}}, \bibinfo {author} {\bibfnamefont {C.}~\bibnamefont {Giunti}}, \bibinfo {author} {\bibfnamefont {Y.}~\bibnamefont {Li}}, \bibinfo {author} {\bibfnamefont {E.}~\bibnamefont {Picciau}}, \bibinfo {author} {\bibfnamefont {C.}~\bibnamefont {Ternes}},\ and\ \bibinfo {author} {\bibfnamefont {Y.}~\bibnamefont {Zhang}},\ }\bibfield  {title} {\bibinfo {title} {{New insights into nuclear physics and weak mixing angle using electroweak probes}},\ }\href@noop {} {\bibfield  {journal} {\bibinfo  {journal} {Phys.Rev.C}\ }\textbf {\bibinfo {volume} {104}},\ \bibinfo {pages} {065502} (\bibinfo {year} {2021}{\natexlab{b}})},\ \Eprint {https://arxiv.org/abs/arXiv:2102.06153} {arXiv:2102.06153 [hep-ph]} \BibitemShut {NoStop}%
\bibitem [{\citenamefont {Cadeddu}\ \emph {et~al.}(2018{\natexlab{b}})\citenamefont {Cadeddu}, \citenamefont {Giunti}, \citenamefont {Kouzakov}, \citenamefont {Li}, \citenamefont {Studenikin},\ and\ \citenamefont {Zhang}}]{Cadeddu:2018dux}%
  \BibitemOpen
  \bibfield  {author} {\bibinfo {author} {\bibfnamefont {M.}~\bibnamefont {Cadeddu}}, \bibinfo {author} {\bibfnamefont {C.}~\bibnamefont {Giunti}}, \bibinfo {author} {\bibfnamefont {K.}~\bibnamefont {Kouzakov}}, \bibinfo {author} {\bibfnamefont {Y.~F.}\ \bibnamefont {Li}}, \bibinfo {author} {\bibfnamefont {A.}~\bibnamefont {Studenikin}},\ and\ \bibinfo {author} {\bibfnamefont {Y.~Y.}\ \bibnamefont {Zhang}},\ }\bibfield  {title} {\bibinfo {title} {{Neutrino Charge Radii from COHERENT Elastic Neutrino-Nucleus Scattering}},\ }\href@noop {} {\bibfield  {journal} {\bibinfo  {journal} {Phys.Rev.}\ }\textbf {\bibinfo {volume} {D98}},\ \bibinfo {pages} {113010} (\bibinfo {year} {2018}{\natexlab{b}})},\ \Eprint {https://arxiv.org/abs/arXiv:1810.05606} {arXiv:1810.05606 [hep-ph]} \BibitemShut {NoStop}%
\bibitem [{\citenamefont {Cadeddu}\ \emph {et~al.}(2020{\natexlab{a}})\citenamefont {Cadeddu}, \citenamefont {Dordei}, \citenamefont {Giunti}, \citenamefont {Li}, \citenamefont {Picciau},\ and\ \citenamefont {Zhang}}]{Cadeddu:2020lky}%
  \BibitemOpen
  \bibfield  {author} {\bibinfo {author} {\bibfnamefont {M.}~\bibnamefont {Cadeddu}}, \bibinfo {author} {\bibfnamefont {F.}~\bibnamefont {Dordei}}, \bibinfo {author} {\bibfnamefont {C.}~\bibnamefont {Giunti}}, \bibinfo {author} {\bibfnamefont {Y.}~\bibnamefont {Li}}, \bibinfo {author} {\bibfnamefont {E.}~\bibnamefont {Picciau}},\ and\ \bibinfo {author} {\bibfnamefont {Y.}~\bibnamefont {Zhang}},\ }\bibfield  {title} {\bibinfo {title} {{Physics results from the first COHERENT observation of CE$\nu$NS in argon and their combination with cesium-iodide data}},\ }\href@noop {} {\bibfield  {journal} {\bibinfo  {journal} {Phys.Rev.}\ }\textbf {\bibinfo {volume} {D102}},\ \bibinfo {pages} {015030} (\bibinfo {year} {2020}{\natexlab{a}})},\ \Eprint {https://arxiv.org/abs/arXiv:2005.01645} {arXiv:2005.01645 [hep-ph]} \BibitemShut {NoStop}%
\bibitem [{\citenamefont {Aristizabal~Sierra}\ \emph {et~al.}(2018)\citenamefont {Aristizabal~Sierra}, \citenamefont {De~Romeri},\ and\ \citenamefont {Rojas}}]{AristizabalSierra:2018eqm}%
  \BibitemOpen
  \bibfield  {author} {\bibinfo {author} {\bibfnamefont {D.}~\bibnamefont {Aristizabal~Sierra}}, \bibinfo {author} {\bibfnamefont {V.}~\bibnamefont {De~Romeri}},\ and\ \bibinfo {author} {\bibfnamefont {N.}~\bibnamefont {Rojas}},\ }\bibfield  {title} {\bibinfo {title} {{COHERENT analysis of neutrino generalized interactions}},\ }\href@noop {} {\bibfield  {journal} {\bibinfo  {journal} {Phys.Rev.}\ }\textbf {\bibinfo {volume} {D98}},\ \bibinfo {pages} {075018} (\bibinfo {year} {2018})},\ \Eprint {https://arxiv.org/abs/arXiv:1806.07424} {arXiv:1806.07424 [hep-ph]} \BibitemShut {NoStop}%
\bibitem [{\citenamefont {Cadeddu}\ \emph {et~al.}(2020{\natexlab{b}})\citenamefont {Cadeddu}, \citenamefont {Dordei}, \citenamefont {Giunti}, \citenamefont {Li},\ and\ \citenamefont {Zhang}}]{Cadeddu:2019eta}%
  \BibitemOpen
  \bibfield  {author} {\bibinfo {author} {\bibfnamefont {M.}~\bibnamefont {Cadeddu}}, \bibinfo {author} {\bibfnamefont {F.}~\bibnamefont {Dordei}}, \bibinfo {author} {\bibfnamefont {C.}~\bibnamefont {Giunti}}, \bibinfo {author} {\bibfnamefont {Y.}~\bibnamefont {Li}},\ and\ \bibinfo {author} {\bibfnamefont {Y.}~\bibnamefont {Zhang}},\ }\bibfield  {title} {\bibinfo {title} {{Neutrino, Electroweak and Nuclear Physics from COHERENT Elastic Neutrino-Nucleus Scattering with Refined Quenching Factor}},\ }\href@noop {} {\bibfield  {journal} {\bibinfo  {journal} {Phys.Rev.}\ }\textbf {\bibinfo {volume} {D101}},\ \bibinfo {pages} {033004} (\bibinfo {year} {2020}{\natexlab{b}})},\ \Eprint {https://arxiv.org/abs/arXiv:1908.06045} {arXiv:1908.06045 [hep-ph]} \BibitemShut {NoStop}%
\bibitem [{\citenamefont {Corona}\ \emph {et~al.}(2022)\citenamefont {Corona}, \citenamefont {Cadeddu}, \citenamefont {Cargioli}, \citenamefont {Dordei}, \citenamefont {Giunti}, \citenamefont {Li}, \citenamefont {Picciau}, \citenamefont {Ternes},\ and\ \citenamefont {Zhang}}]{AtzoriCorona:2022moj}%
  \BibitemOpen
  \bibfield  {author} {\bibinfo {author} {\bibfnamefont {M.~A.}\ \bibnamefont {Corona}}, \bibinfo {author} {\bibfnamefont {M.}~\bibnamefont {Cadeddu}}, \bibinfo {author} {\bibfnamefont {N.}~\bibnamefont {Cargioli}}, \bibinfo {author} {\bibfnamefont {F.}~\bibnamefont {Dordei}}, \bibinfo {author} {\bibfnamefont {C.}~\bibnamefont {Giunti}}, \bibinfo {author} {\bibfnamefont {Y.}~\bibnamefont {Li}}, \bibinfo {author} {\bibfnamefont {E.}~\bibnamefont {Picciau}}, \bibinfo {author} {\bibfnamefont {C.~A.}\ \bibnamefont {Ternes}},\ and\ \bibinfo {author} {\bibfnamefont {Y.}~\bibnamefont {Zhang}},\ }\bibfield  {title} {\bibinfo {title} {{Probing light mediators and $(g-2)_\mu$ through detection of coherent elastic neutrino nucleus scattering at COHERENT}},\ }\href@noop {} {\bibfield  {journal} {\bibinfo  {journal} {JHEP}\ }\textbf {\bibinfo {volume} {05}},\ \bibinfo {pages} {109}},\ \Eprint {https://arxiv.org/abs/arXiv:2202.11002} {arXiv:2202.11002 [hep-ph]} \BibitemShut {NoStop}%
\bibitem [{\citenamefont {Denton}\ \emph {et~al.}(2018)\citenamefont {Denton}, \citenamefont {Farzan},\ and\ \citenamefont {Shoemaker}}]{Denton:2018xmq}%
  \BibitemOpen
  \bibfield  {author} {\bibinfo {author} {\bibfnamefont {P.~B.}\ \bibnamefont {Denton}}, \bibinfo {author} {\bibfnamefont {Y.}~\bibnamefont {Farzan}},\ and\ \bibinfo {author} {\bibfnamefont {I.~M.}\ \bibnamefont {Shoemaker}},\ }\bibfield  {title} {\bibinfo {title} {{Testing large non-standard neutrino interactions with arbitrary mediator mass after COHERENT data}},\ }\href {https://doi.org/10.1007/JHEP07(2018)037} {\bibfield  {journal} {\bibinfo  {journal} {JHEP}\ }\textbf {\bibinfo {volume} {07}},\ \bibinfo {pages} {037}},\ \Eprint {https://arxiv.org/abs/1804.03660} {arXiv:1804.03660 [hep-ph]} \BibitemShut {NoStop}%
\bibitem [{\citenamefont {Denton}\ and\ \citenamefont {Gehrlein}(2021)}]{Denton:2020hop}%
  \BibitemOpen
  \bibfield  {author} {\bibinfo {author} {\bibfnamefont {P.~B.}\ \bibnamefont {Denton}}\ and\ \bibinfo {author} {\bibfnamefont {J.}~\bibnamefont {Gehrlein}},\ }\bibfield  {title} {\bibinfo {title} {{A Statistical Analysis of the COHERENT Data and Applications to New Physics}},\ }\href {https://doi.org/10.1007/JHEP04(2021)266} {\bibfield  {journal} {\bibinfo  {journal} {JHEP}\ }\textbf {\bibinfo {volume} {04}},\ \bibinfo {pages} {266}},\ \Eprint {https://arxiv.org/abs/2008.06062} {arXiv:2008.06062 [hep-ph]} \BibitemShut {NoStop}%
\bibitem [{\citenamefont {Co'}\ \emph {et~al.}(2020)\citenamefont {Co'}, \citenamefont {Anguiano},\ and\ \citenamefont {Lallena}}]{Co:2020gwl}%
  \BibitemOpen
  \bibfield  {author} {\bibinfo {author} {\bibfnamefont {G.}~\bibnamefont {Co'}}, \bibinfo {author} {\bibfnamefont {M.}~\bibnamefont {Anguiano}},\ and\ \bibinfo {author} {\bibfnamefont {A.~M.}\ \bibnamefont {Lallena}},\ }\bibfield  {title} {\bibinfo {title} {{Nuclear structure uncertainties in coherent elastic neutrino-nucleus scattering}},\ }\href {https://doi.org/10.1088/1475-7516/2020/04/044} {\bibfield  {journal} {\bibinfo  {journal} {JCAP}\ }\textbf {\bibinfo {volume} {04}},\ \bibinfo {pages} {044}},\ \Eprint {https://arxiv.org/abs/2001.04684} {arXiv:2001.04684 [nucl-th]} \BibitemShut {NoStop}%
\bibitem [{\citenamefont {Bonet}\ \emph {et~al.}(2021{\natexlab{b}})\citenamefont {Bonet} \emph {et~al.}}]{Bonet:2020ntx}%
  \BibitemOpen
  \bibfield  {author} {\bibinfo {author} {\bibfnamefont {H.}~\bibnamefont {Bonet}} \emph {et~al.},\ }\bibfield  {title} {\bibinfo {title} {{Large-size sub-keV sensitive germanium detectors for the CONUS experiment}},\ }\href {https://doi.org/10.1140/epjc/s10052-021-09038-3} {\bibfield  {journal} {\bibinfo  {journal} {Eur. Phys. J. C}\ }\textbf {\bibinfo {volume} {81}},\ \bibinfo {pages} {267} (\bibinfo {year} {2021}{\natexlab{b}})},\ \Eprint {https://arxiv.org/abs/2010.11241} {arXiv:2010.11241 [physics.ins-det]} \BibitemShut {NoStop}%
\bibitem [{\citenamefont {Augier}\ \emph {et~al.}(2023)\citenamefont {Augier} \emph {et~al.}}]{Ricochet:2022pzj}%
  \BibitemOpen
  \bibfield  {author} {\bibinfo {author} {\bibfnamefont {C.}~\bibnamefont {Augier}} \emph {et~al.} (\bibinfo {collaboration} {Ricochet}),\ }\bibfield  {title} {\bibinfo {title} {{Fast neutron background characterization of the future Ricochet experiment at the ILL research nuclear reactor}},\ }\href {https://doi.org/10.1140/epjc/s10052-022-11150-x} {\bibfield  {journal} {\bibinfo  {journal} {Eur. Phys. J. C}\ }\textbf {\bibinfo {volume} {83}},\ \bibinfo {pages} {20} (\bibinfo {year} {2023})},\ \Eprint {https://arxiv.org/abs/2208.01760} {arXiv:2208.01760 [astro-ph.IM]} \BibitemShut {NoStop}%
\bibitem [{\citenamefont {Atzori~Corona}\ \emph {et~al.}(2022)\citenamefont {Atzori~Corona}, \citenamefont {Cadeddu}, \citenamefont {Cargioli}, \citenamefont {Dordei}, \citenamefont {Giunti}, \citenamefont {Li}, \citenamefont {Ternes},\ and\ \citenamefont {Zhang}}]{AtzoriCorona:2022qrf}%
  \BibitemOpen
  \bibfield  {author} {\bibinfo {author} {\bibfnamefont {M.}~\bibnamefont {Atzori~Corona}}, \bibinfo {author} {\bibfnamefont {M.}~\bibnamefont {Cadeddu}}, \bibinfo {author} {\bibfnamefont {N.}~\bibnamefont {Cargioli}}, \bibinfo {author} {\bibfnamefont {F.}~\bibnamefont {Dordei}}, \bibinfo {author} {\bibfnamefont {C.}~\bibnamefont {Giunti}}, \bibinfo {author} {\bibfnamefont {Y.~F.}\ \bibnamefont {Li}}, \bibinfo {author} {\bibfnamefont {C.~A.}\ \bibnamefont {Ternes}},\ and\ \bibinfo {author} {\bibfnamefont {Y.~Y.}\ \bibnamefont {Zhang}},\ }\bibfield  {title} {\bibinfo {title} {{Impact of the Dresden-II and COHERENT neutrino scattering data on neutrino electromagnetic properties and electroweak physics}},\ }\href {https://doi.org/10.1007/JHEP09(2022)164} {\bibfield  {journal} {\bibinfo  {journal} {JHEP}\ }\textbf {\bibinfo {volume} {09}},\ \bibinfo {pages} {164}},\ \Eprint {https://arxiv.org/abs/2205.09484} {arXiv:2205.09484 [hep-ph]} \BibitemShut {NoStop}%
\bibitem [{\citenamefont {Coloma}\ \emph {et~al.}(2022)\citenamefont {Coloma}, \citenamefont {Esteban}, \citenamefont {Gonzalez-Garcia}, \citenamefont {Larizgoitia}, \citenamefont {Monrabal},\ and\ \citenamefont {Palomares-Ruiz}}]{Coloma:2022avw}%
  \BibitemOpen
  \bibfield  {author} {\bibinfo {author} {\bibfnamefont {P.}~\bibnamefont {Coloma}}, \bibinfo {author} {\bibfnamefont {I.}~\bibnamefont {Esteban}}, \bibinfo {author} {\bibfnamefont {M.~C.}\ \bibnamefont {Gonzalez-Garcia}}, \bibinfo {author} {\bibfnamefont {L.}~\bibnamefont {Larizgoitia}}, \bibinfo {author} {\bibfnamefont {F.}~\bibnamefont {Monrabal}},\ and\ \bibinfo {author} {\bibfnamefont {S.}~\bibnamefont {Palomares-Ruiz}},\ }\bibfield  {title} {\bibinfo {title} {{Bounds on new physics with data of the Dresden-II reactor experiment and COHERENT}},\ }\href {https://doi.org/10.1007/JHEP05(2022)037} {\bibfield  {journal} {\bibinfo  {journal} {JHEP}\ }\textbf {\bibinfo {volume} {05}},\ \bibinfo {pages} {037}},\ \Eprint {https://arxiv.org/abs/2202.10829} {arXiv:2202.10829 [hep-ph]} \BibitemShut {NoStop}%
\bibitem [{\citenamefont {Majumdar}\ \emph {et~al.}(2022)\citenamefont {Majumdar}, \citenamefont {Papoulias}, \citenamefont {Srivastava},\ and\ \citenamefont {Valle}}]{Majumdar:2022nby}%
  \BibitemOpen
  \bibfield  {author} {\bibinfo {author} {\bibfnamefont {A.}~\bibnamefont {Majumdar}}, \bibinfo {author} {\bibfnamefont {D.~K.}\ \bibnamefont {Papoulias}}, \bibinfo {author} {\bibfnamefont {R.}~\bibnamefont {Srivastava}},\ and\ \bibinfo {author} {\bibfnamefont {J.~W.~F.}\ \bibnamefont {Valle}},\ }\bibfield  {title} {\bibinfo {title} {{Physics implications of recent Dresden-II reactor data}},\ }\href {https://doi.org/10.1103/PhysRevD.106.093010} {\bibfield  {journal} {\bibinfo  {journal} {Phys. Rev. D}\ }\textbf {\bibinfo {volume} {106}},\ \bibinfo {pages} {093010} (\bibinfo {year} {2022})},\ \Eprint {https://arxiv.org/abs/2208.13262} {arXiv:2208.13262 [hep-ph]} \BibitemShut {NoStop}%
\bibitem [{\citenamefont {Denton}\ and\ \citenamefont {Gehrlein}(2022)}]{Denton:2022nol}%
  \BibitemOpen
  \bibfield  {author} {\bibinfo {author} {\bibfnamefont {P.~B.}\ \bibnamefont {Denton}}\ and\ \bibinfo {author} {\bibfnamefont {J.}~\bibnamefont {Gehrlein}},\ }\bibfield  {title} {\bibinfo {title} {{New constraints on the dark side of non-standard interactions from reactor neutrino scattering data}},\ }\href {https://doi.org/10.1103/PhysRevD.106.015022} {\bibfield  {journal} {\bibinfo  {journal} {Phys. Rev. D}\ }\textbf {\bibinfo {volume} {106}},\ \bibinfo {pages} {015022} (\bibinfo {year} {2022})},\ \Eprint {https://arxiv.org/abs/2204.09060} {arXiv:2204.09060 [hep-ph]} \BibitemShut {NoStop}%
\bibitem [{\citenamefont {Collar}\ \emph {et~al.}(2021)\citenamefont {Collar}, \citenamefont {Kavner},\ and\ \citenamefont {Lewis}}]{Collar:2021fcl}%
  \BibitemOpen
  \bibfield  {author} {\bibinfo {author} {\bibfnamefont {J.~I.}\ \bibnamefont {Collar}}, \bibinfo {author} {\bibfnamefont {A.~R.~L.}\ \bibnamefont {Kavner}},\ and\ \bibinfo {author} {\bibfnamefont {C.~M.}\ \bibnamefont {Lewis}},\ }\bibfield  {title} {\bibinfo {title} {{Germanium response to sub-keV nuclear recoils: a multipronged experimental characterization}},\ }\href {https://doi.org/10.1103/PhysRevD.103.122003} {\bibfield  {journal} {\bibinfo  {journal} {Phys. Rev. D}\ }\textbf {\bibinfo {volume} {103}},\ \bibinfo {pages} {122003} (\bibinfo {year} {2021})},\ \Eprint {https://arxiv.org/abs/2102.10089} {arXiv:2102.10089 [nucl-ex]} \BibitemShut {NoStop}%
\bibitem [{\citenamefont {Lindhard}\ \emph {et~al.}(1963)\citenamefont {Lindhard}, \citenamefont {Nielsen}, \citenamefont {Scharff},\ and\ \citenamefont {Thomsen}}]{Lindhard_theo}%
  \BibitemOpen
  \bibfield  {author} {\bibinfo {author} {\bibfnamefont {J.}~\bibnamefont {Lindhard}}, \bibinfo {author} {\bibfnamefont {V.}~\bibnamefont {Nielsen}}, \bibinfo {author} {\bibfnamefont {M.}~\bibnamefont {Scharff}},\ and\ \bibinfo {author} {\bibfnamefont {P.~V.}\ \bibnamefont {Thomsen}},\ }\bibfield  {title} {\bibinfo {title} {Integral equations governing radiation effects. (notes on atomic collisions, iii)},\ }\bibfield  {journal} {\bibinfo  {journal} {Kgl. Danske Videnskab., Selskab. Mat. Fys. Medd.}\ }\textbf {\bibinfo {volume} {33, 10}},\ \href {https://www.osti.gov/biblio/4701226} {} (\bibinfo {year} {1963})\BibitemShut {NoStop}%
\bibitem [{\citenamefont {Albakry}\ \emph {et~al.}(2022)\citenamefont {Albakry} \emph {et~al.}}]{SuperCDMS:2022nlc}%
  \BibitemOpen
  \bibfield  {author} {\bibinfo {author} {\bibfnamefont {M.~F.}\ \bibnamefont {Albakry}} \emph {et~al.} (\bibinfo {collaboration} {SuperCDMS}),\ }\bibfield  {title} {\bibinfo {title} {{Ionization yield measurement in a germanium CDMSlite detector using photo-neutron sources}},\ }\href {https://doi.org/10.1103/PhysRevD.105.122002} {\bibfield  {journal} {\bibinfo  {journal} {Phys. Rev. D}\ }\textbf {\bibinfo {volume} {105}},\ \bibinfo {pages} {122002} (\bibinfo {year} {2022})},\ \Eprint {https://arxiv.org/abs/2202.07043} {arXiv:2202.07043 [physics.ins-det]} \BibitemShut {NoStop}%
\bibitem [{\citenamefont {Bonhomme}\ \emph {et~al.}(2022)\citenamefont {Bonhomme} \emph {et~al.}}]{Bonhomme:2022lcz}%
  \BibitemOpen
  \bibfield  {author} {\bibinfo {author} {\bibfnamefont {A.}~\bibnamefont {Bonhomme}} \emph {et~al.},\ }\bibfield  {title} {\bibinfo {title} {{Direct measurement of the ionization quenching factor of nuclear recoils in germanium in the keV energy range}},\ }\href {https://doi.org/10.1140/epjc/s10052-022-10768-1} {\bibfield  {journal} {\bibinfo  {journal} {Eur. Phys. J. C}\ }\textbf {\bibinfo {volume} {82}},\ \bibinfo {pages} {815} (\bibinfo {year} {2022})},\ \Eprint {https://arxiv.org/abs/2202.03754} {arXiv:2202.03754 [physics.ins-det]} \BibitemShut {NoStop}%
\bibitem [{\citenamefont {Xu}\ \emph {et~al.}(2023)\citenamefont {Xu} \emph {et~al.}}]{Xu:2023wev}%
  \BibitemOpen
  \bibfield  {author} {\bibinfo {author} {\bibfnamefont {J.}~\bibnamefont {Xu}} \emph {et~al.},\ }\href@noop {} {\bibinfo {title} {{Search for the Migdal effect in liquid xenon with keV-level nuclear recoils}}} (\bibinfo {year} {2023}),\ \Eprint {https://arxiv.org/abs/2307.12952} {arXiv:2307.12952 [hep-ex]} \BibitemShut {NoStop}%
\bibitem [{\citenamefont {Migdal}(1941)}]{Migdal}%
  \BibitemOpen
  \bibfield  {author} {\bibinfo {author} {\bibfnamefont {A.}~\bibnamefont {Migdal}},\ }\bibfield  {title} {\bibinfo {title} {{Ionization of atoms accompanying $\alpha$- and $\beta$-decay, J. Phys}},\ }\href@noop {} {\bibfield  {journal} {\bibinfo  {journal} {USSR}\ }\textbf {\bibinfo {volume} {4}},\ \bibinfo {pages} {449} (\bibinfo {year} {1941})}\BibitemShut {NoStop}%
\bibitem [{\citenamefont {Ibe}\ \emph {et~al.}(2018)\citenamefont {Ibe}, \citenamefont {Nakano}, \citenamefont {Shoji},\ and\ \citenamefont {Suzuki}}]{Ibe:2017yqa}%
  \BibitemOpen
  \bibfield  {author} {\bibinfo {author} {\bibfnamefont {M.}~\bibnamefont {Ibe}}, \bibinfo {author} {\bibfnamefont {W.}~\bibnamefont {Nakano}}, \bibinfo {author} {\bibfnamefont {Y.}~\bibnamefont {Shoji}},\ and\ \bibinfo {author} {\bibfnamefont {K.}~\bibnamefont {Suzuki}},\ }\bibfield  {title} {\bibinfo {title} {{Migdal Effect in Dark Matter Direct Detection Experiments}},\ }\href {https://doi.org/10.1007/JHEP03(2018)194} {\bibfield  {journal} {\bibinfo  {journal} {JHEP}\ }\textbf {\bibinfo {volume} {03}},\ \bibinfo {pages} {194}},\ \Eprint {https://arxiv.org/abs/1707.07258} {arXiv:1707.07258 [hep-ph]} \BibitemShut {NoStop}%
\bibitem [{\citenamefont {Essig}\ \emph {et~al.}(2020)\citenamefont {Essig}, \citenamefont {Pradler}, \citenamefont {Sholapurkar},\ and\ \citenamefont {Yu}}]{Essig:2019xkx}%
  \BibitemOpen
  \bibfield  {author} {\bibinfo {author} {\bibfnamefont {R.}~\bibnamefont {Essig}}, \bibinfo {author} {\bibfnamefont {J.}~\bibnamefont {Pradler}}, \bibinfo {author} {\bibfnamefont {M.}~\bibnamefont {Sholapurkar}},\ and\ \bibinfo {author} {\bibfnamefont {T.-T.}\ \bibnamefont {Yu}},\ }\bibfield  {title} {\bibinfo {title} {{Relation between the Migdal Effect and Dark Matter-Electron Scattering in Isolated Atoms and Semiconductors}},\ }\href {https://doi.org/10.1103/PhysRevLett.124.021801} {\bibfield  {journal} {\bibinfo  {journal} {Phys. Rev. Lett.}\ }\textbf {\bibinfo {volume} {124}},\ \bibinfo {pages} {021801} (\bibinfo {year} {2020})},\ \Eprint {https://arxiv.org/abs/1908.10881} {arXiv:1908.10881 [hep-ph]} \BibitemShut {NoStop}%
\bibitem [{\citenamefont {Grilli~di Cortona}\ \emph {et~al.}(2020)\citenamefont {Grilli~di Cortona}, \citenamefont {Messina},\ and\ \citenamefont {Piacentini}}]{GrillidiCortona:2020owp}%
  \BibitemOpen
  \bibfield  {author} {\bibinfo {author} {\bibfnamefont {G.}~\bibnamefont {Grilli~di Cortona}}, \bibinfo {author} {\bibfnamefont {A.}~\bibnamefont {Messina}},\ and\ \bibinfo {author} {\bibfnamefont {S.}~\bibnamefont {Piacentini}},\ }\bibfield  {title} {\bibinfo {title} {{Migdal effect and photon Bremsstrahlung: improving the sensitivity to light dark matter of liquid argon experiments}},\ }\href {https://doi.org/10.1007/JHEP11(2020)034} {\bibfield  {journal} {\bibinfo  {journal} {JHEP}\ }\textbf {\bibinfo {volume} {11}},\ \bibinfo {pages} {034}},\ \Eprint {https://arxiv.org/abs/2006.02453} {arXiv:2006.02453 [hep-ph]} \BibitemShut {NoStop}%
\bibitem [{\citenamefont {Liu}\ \emph {et~al.}(2020)\citenamefont {Liu}, \citenamefont {Wu}, \citenamefont {Chi},\ and\ \citenamefont {Chen}}]{Liu:2020pat}%
  \BibitemOpen
  \bibfield  {author} {\bibinfo {author} {\bibfnamefont {C.~P.}\ \bibnamefont {Liu}}, \bibinfo {author} {\bibfnamefont {C.-P.}\ \bibnamefont {Wu}}, \bibinfo {author} {\bibfnamefont {H.-C.}\ \bibnamefont {Chi}},\ and\ \bibinfo {author} {\bibfnamefont {J.-W.}\ \bibnamefont {Chen}},\ }\bibfield  {title} {\bibinfo {title} {{Model-independent determination of the Migdal effect via photoabsorption}},\ }\href {https://doi.org/10.1103/PhysRevD.102.121303} {\bibfield  {journal} {\bibinfo  {journal} {Phys. Rev. D}\ }\textbf {\bibinfo {volume} {102}},\ \bibinfo {pages} {121303} (\bibinfo {year} {2020})},\ \Eprint {https://arxiv.org/abs/2007.10965} {arXiv:2007.10965 [hep-ph]} \BibitemShut {NoStop}%
\bibitem [{\citenamefont {Agnes}\ \emph {et~al.}(2023)\citenamefont {Agnes} \emph {et~al.}}]{DarkSide:2022dhx}%
  \BibitemOpen
  \bibfield  {author} {\bibinfo {author} {\bibfnamefont {P.}~\bibnamefont {Agnes}} \emph {et~al.} (\bibinfo {collaboration} {DarkSide}),\ }\bibfield  {title} {\bibinfo {title} {{Search for Dark-Matter\textendash{}Nucleon Interactions via Migdal Effect with DarkSide-50}},\ }\href {https://doi.org/10.1103/PhysRevLett.130.101001} {\bibfield  {journal} {\bibinfo  {journal} {Phys. Rev. Lett.}\ }\textbf {\bibinfo {volume} {130}},\ \bibinfo {pages} {101001} (\bibinfo {year} {2023})},\ \Eprint {https://arxiv.org/abs/2207.11967} {arXiv:2207.11967 [hep-ex]} \BibitemShut {NoStop}%
\bibitem [{\citenamefont {Aprile}\ \emph {et~al.}(2019)\citenamefont {Aprile} \emph {et~al.}}]{XENON:2019zpr}%
  \BibitemOpen
  \bibfield  {author} {\bibinfo {author} {\bibfnamefont {E.}~\bibnamefont {Aprile}} \emph {et~al.} (\bibinfo {collaboration} {XENON}),\ }\bibfield  {title} {\bibinfo {title} {{Search for Light Dark Matter Interactions Enhanced by the Migdal Effect or Bremsstrahlung in XENON1T}},\ }\href {https://doi.org/10.1103/PhysRevLett.123.241803} {\bibfield  {journal} {\bibinfo  {journal} {Phys. Rev. Lett.}\ }\textbf {\bibinfo {volume} {123}},\ \bibinfo {pages} {241803} (\bibinfo {year} {2019})},\ \Eprint {https://arxiv.org/abs/1907.12771} {arXiv:1907.12771 [hep-ex]} \BibitemShut {NoStop}%
\bibitem [{\citenamefont {Albakry}\ \emph {et~al.}(2023)\citenamefont {Albakry} \emph {et~al.}}]{SuperCDMS:2023sql}%
  \BibitemOpen
  \bibfield  {author} {\bibinfo {author} {\bibfnamefont {M.~F.}\ \bibnamefont {Albakry}} \emph {et~al.} (\bibinfo {collaboration} {SuperCDMS}),\ }\bibfield  {title} {\bibinfo {title} {{Search for low-mass dark matter via bremsstrahlung radiation and the Migdal effect in SuperCDMS}},\ }\href {https://doi.org/10.1103/PhysRevD.107.112013} {\bibfield  {journal} {\bibinfo  {journal} {Phys. Rev. D}\ }\textbf {\bibinfo {volume} {107}},\ \bibinfo {pages} {112013} (\bibinfo {year} {2023})},\ \Eprint {https://arxiv.org/abs/2302.09115} {arXiv:2302.09115 [hep-ex]} \BibitemShut {NoStop}%
\bibitem [{\citenamefont {Bell}\ \emph {et~al.}(2021)\citenamefont {Bell}, \citenamefont {Dent}, \citenamefont {Dutta}, \citenamefont {Ghosh}, \citenamefont {Kumar},\ and\ \citenamefont {Newstead}}]{Bell:2021zkr}%
  \BibitemOpen
  \bibfield  {author} {\bibinfo {author} {\bibfnamefont {N.~F.}\ \bibnamefont {Bell}}, \bibinfo {author} {\bibfnamefont {J.~B.}\ \bibnamefont {Dent}}, \bibinfo {author} {\bibfnamefont {B.}~\bibnamefont {Dutta}}, \bibinfo {author} {\bibfnamefont {S.}~\bibnamefont {Ghosh}}, \bibinfo {author} {\bibfnamefont {J.}~\bibnamefont {Kumar}},\ and\ \bibinfo {author} {\bibfnamefont {J.~L.}\ \bibnamefont {Newstead}},\ }\bibfield  {title} {\bibinfo {title} {{Low-mass inelastic dark matter direct detection via the Migdal effect}},\ }\href {https://doi.org/10.1103/PhysRevD.104.076013} {\bibfield  {journal} {\bibinfo  {journal} {Phys. Rev. D}\ }\textbf {\bibinfo {volume} {104}},\ \bibinfo {pages} {076013} (\bibinfo {year} {2021})},\ \Eprint {https://arxiv.org/abs/2103.05890} {arXiv:2103.05890 [hep-ph]} \BibitemShut {NoStop}%
\bibitem [{\citenamefont {Bell}\ \emph {et~al.}(2020)\citenamefont {Bell}, \citenamefont {Dent}, \citenamefont {Newstead}, \citenamefont {Sabharwal},\ and\ \citenamefont {Weiler}}]{Bell:2019egg}%
  \BibitemOpen
  \bibfield  {author} {\bibinfo {author} {\bibfnamefont {N.~F.}\ \bibnamefont {Bell}}, \bibinfo {author} {\bibfnamefont {J.~B.}\ \bibnamefont {Dent}}, \bibinfo {author} {\bibfnamefont {J.~L.}\ \bibnamefont {Newstead}}, \bibinfo {author} {\bibfnamefont {S.}~\bibnamefont {Sabharwal}},\ and\ \bibinfo {author} {\bibfnamefont {T.~J.}\ \bibnamefont {Weiler}},\ }\bibfield  {title} {\bibinfo {title} {{Migdal effect and photon bremsstrahlung in effective field theories of dark matter direct detection and coherent elastic neutrino-nucleus scattering}},\ }\href {https://doi.org/10.1103/PhysRevD.101.015012} {\bibfield  {journal} {\bibinfo  {journal} {Phys. Rev. D}\ }\textbf {\bibinfo {volume} {101}},\ \bibinfo {pages} {015012} (\bibinfo {year} {2020})},\ \Eprint {https://arxiv.org/abs/1905.00046} {arXiv:1905.00046 [hep-ph]} \BibitemShut {NoStop}%
\bibitem [{\citenamefont {Bell}\ \emph {et~al.}(2022)\citenamefont {Bell}, \citenamefont {Dent}, \citenamefont {Lang}, \citenamefont {Newstead},\ and\ \citenamefont {Ritter}}]{Bell:2021ihi}%
  \BibitemOpen
  \bibfield  {author} {\bibinfo {author} {\bibfnamefont {N.~F.}\ \bibnamefont {Bell}}, \bibinfo {author} {\bibfnamefont {J.~B.}\ \bibnamefont {Dent}}, \bibinfo {author} {\bibfnamefont {R.~F.}\ \bibnamefont {Lang}}, \bibinfo {author} {\bibfnamefont {J.~L.}\ \bibnamefont {Newstead}},\ and\ \bibinfo {author} {\bibfnamefont {A.~C.}\ \bibnamefont {Ritter}},\ }\bibfield  {title} {\bibinfo {title} {{Observing the Migdal effect from nuclear recoils of neutral particles with liquid xenon and argon detectors}},\ }\href {https://doi.org/10.1103/PhysRevD.105.096015} {\bibfield  {journal} {\bibinfo  {journal} {Phys. Rev. D}\ }\textbf {\bibinfo {volume} {105}},\ \bibinfo {pages} {096015} (\bibinfo {year} {2022})},\ \Eprint {https://arxiv.org/abs/2112.08514} {arXiv:2112.08514 [hep-ph]} \BibitemShut {NoStop}%
\bibitem [{\citenamefont {Liao}\ \emph {et~al.}(2021)\citenamefont {Liao}, \citenamefont {Liu},\ and\ \citenamefont {Marfatia}}]{Liao:2021yog}%
  \BibitemOpen
  \bibfield  {author} {\bibinfo {author} {\bibfnamefont {J.}~\bibnamefont {Liao}}, \bibinfo {author} {\bibfnamefont {H.}~\bibnamefont {Liu}},\ and\ \bibinfo {author} {\bibfnamefont {D.}~\bibnamefont {Marfatia}},\ }\bibfield  {title} {\bibinfo {title} {{Coherent neutrino scattering and the Migdal effect on the quenching factor}},\ }\href {https://doi.org/10.1103/PhysRevD.104.015005} {\bibfield  {journal} {\bibinfo  {journal} {Phys. Rev. D}\ }\textbf {\bibinfo {volume} {104}},\ \bibinfo {pages} {015005} (\bibinfo {year} {2021})},\ \Eprint {https://arxiv.org/abs/2104.01811} {arXiv:2104.01811 [hep-ph]} \BibitemShut {NoStop}%
\bibitem [{\citenamefont {Liao}\ \emph {et~al.}(2022)\citenamefont {Liao}, \citenamefont {Liu},\ and\ \citenamefont {Marfatia}}]{Liao:2022hno}%
  \BibitemOpen
  \bibfield  {author} {\bibinfo {author} {\bibfnamefont {J.}~\bibnamefont {Liao}}, \bibinfo {author} {\bibfnamefont {H.}~\bibnamefont {Liu}},\ and\ \bibinfo {author} {\bibfnamefont {D.}~\bibnamefont {Marfatia}},\ }\bibfield  {title} {\bibinfo {title} {{Implications of the first evidence for coherent elastic scattering of reactor neutrinos}},\ }\href {https://doi.org/10.1103/PhysRevD.106.L031702} {\bibfield  {journal} {\bibinfo  {journal} {Phys. Rev. D}\ }\textbf {\bibinfo {volume} {106}},\ \bibinfo {pages} {L031702} (\bibinfo {year} {2022})},\ \Eprint {https://arxiv.org/abs/2202.10622} {arXiv:2202.10622 [hep-ph]} \BibitemShut {NoStop}%
\bibitem [{\citenamefont {Sorensen}(2015)}]{Sorensen:2014sla}%
  \BibitemOpen
  \bibfield  {author} {\bibinfo {author} {\bibfnamefont {P.}~\bibnamefont {Sorensen}},\ }\bibfield  {title} {\bibinfo {title} {{Atomic limits in the search for galactic dark matter}},\ }\href {https://doi.org/10.1103/PhysRevD.91.083509} {\bibfield  {journal} {\bibinfo  {journal} {Phys. Rev. D}\ }\textbf {\bibinfo {volume} {91}},\ \bibinfo {pages} {083509} (\bibinfo {year} {2015})},\ \Eprint {https://arxiv.org/abs/1412.3028} {arXiv:1412.3028 [astro-ph.IM]} \BibitemShut {NoStop}%
\bibitem [{\citenamefont {Drukier}\ and\ \citenamefont {Stodolsky}(1984)}]{PhysRevD.30.2295}%
  \BibitemOpen
  \bibfield  {author} {\bibinfo {author} {\bibfnamefont {A.}~\bibnamefont {Drukier}}\ and\ \bibinfo {author} {\bibfnamefont {L.}~\bibnamefont {Stodolsky}},\ }\bibfield  {title} {\bibinfo {title} {Principles and applications of a neutral-current detector for neutrino physics and astronomy},\ }\href {https://doi.org/10.1103/PhysRevD.30.2295} {\bibfield  {journal} {\bibinfo  {journal} {Phys. Rev. D}\ }\textbf {\bibinfo {volume} {30}},\ \bibinfo {pages} {2295} (\bibinfo {year} {1984})}\BibitemShut {NoStop}%
\bibitem [{\citenamefont {Barranco}\ \emph {et~al.}(2005)\citenamefont {Barranco}, \citenamefont {Miranda},\ and\ \citenamefont {Rashba}}]{Barranco:2005yy}%
  \BibitemOpen
  \bibfield  {author} {\bibinfo {author} {\bibfnamefont {J.}~\bibnamefont {Barranco}}, \bibinfo {author} {\bibfnamefont {O.~G.}\ \bibnamefont {Miranda}},\ and\ \bibinfo {author} {\bibfnamefont {T.~I.}\ \bibnamefont {Rashba}},\ }\bibfield  {title} {\bibinfo {title} {{Probing new physics with coherent neutrino scattering off nuclei}},\ }\href {https://doi.org/10.1088/1126-6708/2005/12/021} {\bibfield  {journal} {\bibinfo  {journal} {JHEP}\ }\textbf {\bibinfo {volume} {12}},\ \bibinfo {pages} {021}},\ \Eprint {https://arxiv.org/abs/hep-ph/0508299} {arXiv:hep-ph/0508299} \BibitemShut {NoStop}%
\bibitem [{\citenamefont {Atzori~Corona}\ \emph {et~al.}(2023{\natexlab{a}})\citenamefont {Atzori~Corona}, \citenamefont {Cadeddu}, \citenamefont {Cargioli}, \citenamefont {Dordei}, \citenamefont {Giunti},\ and\ \citenamefont {Masia}}]{AtzoriCorona:2023ktl}%
  \BibitemOpen
  \bibfield  {author} {\bibinfo {author} {\bibfnamefont {M.}~\bibnamefont {Atzori~Corona}}, \bibinfo {author} {\bibfnamefont {M.}~\bibnamefont {Cadeddu}}, \bibinfo {author} {\bibfnamefont {N.}~\bibnamefont {Cargioli}}, \bibinfo {author} {\bibfnamefont {F.}~\bibnamefont {Dordei}}, \bibinfo {author} {\bibfnamefont {C.}~\bibnamefont {Giunti}},\ and\ \bibinfo {author} {\bibfnamefont {G.}~\bibnamefont {Masia}},\ }\bibfield  {title} {\bibinfo {title} {{Nuclear neutron radius and weak mixing angle measurements from latest COHERENT CsI and atomic parity violation Cs data}},\ }\href {https://doi.org/10.1140/epjc/s10052-023-11849-5} {\bibfield  {journal} {\bibinfo  {journal} {Eur. Phys. J. C}\ }\textbf {\bibinfo {volume} {83}},\ \bibinfo {pages} {683} (\bibinfo {year} {2023}{\natexlab{a}})},\ \Eprint {https://arxiv.org/abs/2303.09360} {arXiv:2303.09360 [nucl-ex]} \BibitemShut {NoStop}%
\bibitem [{\citenamefont {Cadeddu}\ \emph {et~al.}(2023)\citenamefont {Cadeddu}, \citenamefont {Dordei},\ and\ \citenamefont {Giunti}}]{Cadeddu:2023tkp}%
  \BibitemOpen
  \bibfield  {author} {\bibinfo {author} {\bibfnamefont {M.}~\bibnamefont {Cadeddu}}, \bibinfo {author} {\bibfnamefont {F.}~\bibnamefont {Dordei}},\ and\ \bibinfo {author} {\bibfnamefont {C.}~\bibnamefont {Giunti}},\ }\href@noop {} {\bibinfo {title} {{A view of Coherent Elastic Neutrino-Nucleus Scattering}}} (\bibinfo {year} {2023}),\ \Eprint {https://arxiv.org/abs/2307.08842} {arXiv:2307.08842 [hep-ph]} \BibitemShut {NoStop}%
\bibitem [{\citenamefont {Berglund}\ and\ \citenamefont {Wieser}(2011)}]{BerglundWieser+2011+397+410}%
  \BibitemOpen
  \bibfield  {author} {\bibinfo {author} {\bibfnamefont {M.}~\bibnamefont {Berglund}}\ and\ \bibinfo {author} {\bibfnamefont {M.~E.}\ \bibnamefont {Wieser}},\ }\bibfield  {title} {\bibinfo {title} {Isotopic compositions of the elements 2009 (iupac technical report)},\ }\href {https://doi.org/doi:10.1351/PAC-REP-10-06-02} {\bibfield  {journal} {\bibinfo  {journal} {Pure and Applied Chemistry}\ }\textbf {\bibinfo {volume} {83}},\ \bibinfo {pages} {397} (\bibinfo {year} {2011})}\BibitemShut {NoStop}%
\bibitem [{\citenamefont {Workman}\ \emph {et~al.}(2022)\citenamefont {Workman} \emph {et~al.}}]{ParticleDataGroup:2022pth}%
  \BibitemOpen
  \bibfield  {author} {\bibinfo {author} {\bibfnamefont {R.~L.}\ \bibnamefont {Workman}} \emph {et~al.} (\bibinfo {collaboration} {Particle Data Group}),\ }\bibfield  {title} {\bibinfo {title} {{Review of Particle Physics}},\ }\href {https://doi.org/10.1093/ptep/ptac097} {\bibfield  {journal} {\bibinfo  {journal} {PTEP}\ }\textbf {\bibinfo {volume} {2022}},\ \bibinfo {pages} {083C01} (\bibinfo {year} {2022})}\BibitemShut {NoStop}%
\bibitem [{\citenamefont {Huber}(2011)}]{Huber:2011wv}%
  \BibitemOpen
  \bibfield  {author} {\bibinfo {author} {\bibfnamefont {P.}~\bibnamefont {Huber}},\ }\bibfield  {title} {\bibinfo {title} {{On the determination of anti-neutrino spectra from nuclear reactors}},\ }\href {https://doi.org/10.1103/PhysRevC.85.029901} {\bibfield  {journal} {\bibinfo  {journal} {Phys. Rev. C}\ }\textbf {\bibinfo {volume} {84}},\ \bibinfo {pages} {024617} (\bibinfo {year} {2011})},\ \bibinfo {note} {[Erratum: Phys.Rev.C 85, 029901 (2012)]},\ \Eprint {https://arxiv.org/abs/1106.0687} {arXiv:1106.0687 [hep-ph]} \BibitemShut {NoStop}%
\bibitem [{\citenamefont {Mueller}\ \emph {et~al.}(2011)\citenamefont {Mueller} \emph {et~al.}}]{Mueller:2011nm}%
  \BibitemOpen
  \bibfield  {author} {\bibinfo {author} {\bibfnamefont {T.~A.}\ \bibnamefont {Mueller}} \emph {et~al.},\ }\bibfield  {title} {\bibinfo {title} {{Improved Predictions of Reactor Antineutrino Spectra}},\ }\href {https://doi.org/10.1103/PhysRevC.83.054615} {\bibfield  {journal} {\bibinfo  {journal} {Phys. Rev. C}\ }\textbf {\bibinfo {volume} {83}},\ \bibinfo {pages} {054615} (\bibinfo {year} {2011})},\ \Eprint {https://arxiv.org/abs/1101.2663} {arXiv:1101.2663 [hep-ex]} \BibitemShut {NoStop}%
\bibitem [{\citenamefont {Vogel}\ and\ \citenamefont {Engel}(1989)}]{PhysRevD.39.3378}%
  \BibitemOpen
  \bibfield  {author} {\bibinfo {author} {\bibfnamefont {P.}~\bibnamefont {Vogel}}\ and\ \bibinfo {author} {\bibfnamefont {J.}~\bibnamefont {Engel}},\ }\bibfield  {title} {\bibinfo {title} {Neutrino electromagnetic form factors},\ }\href {https://doi.org/10.1103/PhysRevD.39.3378} {\bibfield  {journal} {\bibinfo  {journal} {Phys. Rev. D}\ }\textbf {\bibinfo {volume} {39}},\ \bibinfo {pages} {3378} (\bibinfo {year} {1989})}\BibitemShut {NoStop}%
\bibitem [{\citenamefont {Estienne}\ \emph {et~al.}(2019)\citenamefont {Estienne} \emph {et~al.}}]{Estienne:2019ujo}%
  \BibitemOpen
  \bibfield  {author} {\bibinfo {author} {\bibfnamefont {M.}~\bibnamefont {Estienne}} \emph {et~al.},\ }\bibfield  {title} {\bibinfo {title} {{Updated Summation Model: An Improved Agreement with the Daya Bay Antineutrino Fluxes}},\ }\href {https://doi.org/10.1103/PhysRevLett.123.022502} {\bibfield  {journal} {\bibinfo  {journal} {Phys. Rev. Lett.}\ }\textbf {\bibinfo {volume} {123}},\ \bibinfo {pages} {022502} (\bibinfo {year} {2019})},\ \Eprint {https://arxiv.org/abs/1904.09358} {arXiv:1904.09358 [nucl-ex]} \BibitemShut {NoStop}%
\bibitem [{\citenamefont {Kopeikin}\ \emph {et~al.}(2000)\citenamefont {Kopeikin}, \citenamefont {Mikaelyan},\ and\ \citenamefont {Sinev}}]{Kopeikin:1999tc}%
  \BibitemOpen
  \bibfield  {author} {\bibinfo {author} {\bibfnamefont {V.~I.}\ \bibnamefont {Kopeikin}}, \bibinfo {author} {\bibfnamefont {L.~A.}\ \bibnamefont {Mikaelyan}},\ and\ \bibinfo {author} {\bibfnamefont {V.~V.}\ \bibnamefont {Sinev}},\ }\bibfield  {title} {\bibinfo {title} {{Search for the neutrino magnetic moment in the nonequilibrium reactor anti-neutrino energy spectrum}},\ }\href {https://doi.org/10.1134/1.855741} {\bibfield  {journal} {\bibinfo  {journal} {Phys. Atom. Nucl.}\ }\textbf {\bibinfo {volume} {63}},\ \bibinfo {pages} {1012} (\bibinfo {year} {2000})},\ \Eprint {https://arxiv.org/abs/hep-ph/9904384} {arXiv:hep-ph/9904384} \BibitemShut {NoStop}%
\bibitem [{\citenamefont {Kopeikin}(2012)}]{Kopeikin:2012zz}%
  \BibitemOpen
  \bibfield  {author} {\bibinfo {author} {\bibfnamefont {V.~I.}\ \bibnamefont {Kopeikin}},\ }\bibfield  {title} {\bibinfo {title} {{Flux and spectrum of reactor antineutrinos}},\ }\href {https://doi.org/10.1134/S1063778812020123} {\bibfield  {journal} {\bibinfo  {journal} {Phys. Atom. Nucl.}\ }\textbf {\bibinfo {volume} {75}},\ \bibinfo {pages} {143} (\bibinfo {year} {2012})}\BibitemShut {NoStop}%
\bibitem [{\citenamefont {Cox}\ \emph {et~al.}(2023)\citenamefont {Cox}, \citenamefont {Dolan}, \citenamefont {McCabe},\ and\ \citenamefont {Quiney}}]{Cox:2022ekg}%
  \BibitemOpen
  \bibfield  {author} {\bibinfo {author} {\bibfnamefont {P.}~\bibnamefont {Cox}}, \bibinfo {author} {\bibfnamefont {M.~J.}\ \bibnamefont {Dolan}}, \bibinfo {author} {\bibfnamefont {C.}~\bibnamefont {McCabe}},\ and\ \bibinfo {author} {\bibfnamefont {H.~M.}\ \bibnamefont {Quiney}},\ }\bibfield  {title} {\bibinfo {title} {{Precise predictions and new insights for atomic ionization from the Migdal effect}},\ }\href {https://doi.org/10.1103/PhysRevD.107.035032} {\bibfield  {journal} {\bibinfo  {journal} {Phys. Rev. D}\ }\textbf {\bibinfo {volume} {107}},\ \bibinfo {pages} {035032} (\bibinfo {year} {2023})},\ \Eprint {https://arxiv.org/abs/2208.12222} {arXiv:2208.12222 [hep-ph]} \BibitemShut {NoStop}%
\bibitem [{\citenamefont {Henke}\ \emph {et~al.}(1993)\citenamefont {Henke}, \citenamefont {Gullikson},\ and\ \citenamefont {Davis}}]{HENKE1993181}%
  \BibitemOpen
  \bibfield  {author} {\bibinfo {author} {\bibfnamefont {B.}~\bibnamefont {Henke}}, \bibinfo {author} {\bibfnamefont {E.}~\bibnamefont {Gullikson}},\ and\ \bibinfo {author} {\bibfnamefont {J.}~\bibnamefont {Davis}},\ }\bibfield  {title} {\bibinfo {title} {X-ray interactions: Photoabsorption, scattering, transmission, and reflection at e = 50-30,000 ev, z = 1-92},\ }\href {https://doi.org/https://doi.org/10.1006/adnd.1993.1013} {\bibfield  {journal} {\bibinfo  {journal} {Atomic Data and Nuclear Data Tables}\ }\textbf {\bibinfo {volume} {54}},\ \bibinfo {pages} {181} (\bibinfo {year} {1993})}\BibitemShut {NoStop}%
\bibitem [{\citenamefont {Chen}\ \emph {et~al.}(2015)\citenamefont {Chen}, \citenamefont {Chi}, \citenamefont {Huang}, \citenamefont {Li}, \citenamefont {Liu}, \citenamefont {Singh}, \citenamefont {Wong}, \citenamefont {Wu},\ and\ \citenamefont {Wu}}]{Chen:2014ypv}%
  \BibitemOpen
  \bibfield  {author} {\bibinfo {author} {\bibfnamefont {J.-W.}\ \bibnamefont {Chen}}, \bibinfo {author} {\bibfnamefont {H.-C.}\ \bibnamefont {Chi}}, \bibinfo {author} {\bibfnamefont {K.-N.}\ \bibnamefont {Huang}}, \bibinfo {author} {\bibfnamefont {H.-B.}\ \bibnamefont {Li}}, \bibinfo {author} {\bibfnamefont {C.~P.}\ \bibnamefont {Liu}}, \bibinfo {author} {\bibfnamefont {L.}~\bibnamefont {Singh}}, \bibinfo {author} {\bibfnamefont {H.~T.}\ \bibnamefont {Wong}}, \bibinfo {author} {\bibfnamefont {C.-L.}\ \bibnamefont {Wu}},\ and\ \bibinfo {author} {\bibfnamefont {C.-P.}\ \bibnamefont {Wu}},\ }\bibfield  {title} {\bibinfo {title} {{Constraining neutrino electromagnetic properties by germanium detectors}},\ }\href {https://doi.org/10.1103/PhysRevD.91.013005} {\bibfield  {journal} {\bibinfo  {journal} {Phys. Rev. D}\ }\textbf {\bibinfo {volume} {91}},\ \bibinfo {pages} {013005} (\bibinfo {year} {2015})},\ \Eprint {https://arxiv.org/abs/1411.0574} {arXiv:1411.0574 [hep-ph]} \BibitemShut {NoStop}%
\bibitem [{\citenamefont {Knapen}\ \emph {et~al.}(2021)\citenamefont {Knapen}, \citenamefont {Kozaczuk},\ and\ \citenamefont {Lin}}]{Knapen:2020aky}%
  \BibitemOpen
  \bibfield  {author} {\bibinfo {author} {\bibfnamefont {S.}~\bibnamefont {Knapen}}, \bibinfo {author} {\bibfnamefont {J.}~\bibnamefont {Kozaczuk}},\ and\ \bibinfo {author} {\bibfnamefont {T.}~\bibnamefont {Lin}},\ }\bibfield  {title} {\bibinfo {title} {{Migdal Effect in Semiconductors}},\ }\href {https://doi.org/10.1103/PhysRevLett.127.081805} {\bibfield  {journal} {\bibinfo  {journal} {Phys. Rev. Lett.}\ }\textbf {\bibinfo {volume} {127}},\ \bibinfo {pages} {081805} (\bibinfo {year} {2021})},\ \Eprint {https://arxiv.org/abs/2011.09496} {arXiv:2011.09496 [hep-ph]} \BibitemShut {NoStop}%
\bibitem [{\citenamefont {Knapen}\ \emph {et~al.}(2022)\citenamefont {Knapen}, \citenamefont {Kozaczuk},\ and\ \citenamefont {Lin}}]{Knapen:2021bwg}%
  \BibitemOpen
  \bibfield  {author} {\bibinfo {author} {\bibfnamefont {S.}~\bibnamefont {Knapen}}, \bibinfo {author} {\bibfnamefont {J.}~\bibnamefont {Kozaczuk}},\ and\ \bibinfo {author} {\bibfnamefont {T.}~\bibnamefont {Lin}},\ }\bibfield  {title} {\bibinfo {title} {{python package for dark matter scattering in dielectric targets}},\ }\href {https://doi.org/10.1103/PhysRevD.105.015014} {\bibfield  {journal} {\bibinfo  {journal} {Phys. Rev. D}\ }\textbf {\bibinfo {volume} {105}},\ \bibinfo {pages} {015014} (\bibinfo {year} {2022})},\ \Eprint {https://arxiv.org/abs/2104.12786} {arXiv:2104.12786 [hep-ph]} \BibitemShut {NoStop}%
\bibitem [{\citenamefont {Adams}\ \emph {et~al.}(2023)\citenamefont {Adams}, \citenamefont {Baxter}, \citenamefont {Day}, \citenamefont {Essig},\ and\ \citenamefont {Kahn}}]{Adams:2022zvg}%
  \BibitemOpen
  \bibfield  {author} {\bibinfo {author} {\bibfnamefont {D.}~\bibnamefont {Adams}}, \bibinfo {author} {\bibfnamefont {D.}~\bibnamefont {Baxter}}, \bibinfo {author} {\bibfnamefont {H.}~\bibnamefont {Day}}, \bibinfo {author} {\bibfnamefont {R.}~\bibnamefont {Essig}},\ and\ \bibinfo {author} {\bibfnamefont {Y.}~\bibnamefont {Kahn}},\ }\bibfield  {title} {\bibinfo {title} {{Measuring the Migdal effect in semiconductors for dark matter detection}},\ }\href {https://doi.org/10.1103/PhysRevD.107.L041303} {\bibfield  {journal} {\bibinfo  {journal} {Phys. Rev. D}\ }\textbf {\bibinfo {volume} {107}},\ \bibinfo {pages} {L041303} (\bibinfo {year} {2023})},\ \Eprint {https://arxiv.org/abs/2210.04917} {arXiv:2210.04917 [hep-ph]} \BibitemShut {NoStop}%
\bibitem [{\citenamefont {Liang}\ \emph {et~al.}(2022)\citenamefont {Liang}, \citenamefont {Mo}, \citenamefont {Zheng},\ and\ \citenamefont {Zhang}}]{Liang:2022xbu}%
  \BibitemOpen
  \bibfield  {author} {\bibinfo {author} {\bibfnamefont {Z.-L.}\ \bibnamefont {Liang}}, \bibinfo {author} {\bibfnamefont {C.}~\bibnamefont {Mo}}, \bibinfo {author} {\bibfnamefont {F.}~\bibnamefont {Zheng}},\ and\ \bibinfo {author} {\bibfnamefont {P.}~\bibnamefont {Zhang}},\ }\bibfield  {title} {\bibinfo {title} {{Phonon-mediated Migdal effect in semiconductor detectors}},\ }\href {https://doi.org/10.1103/PhysRevD.106.043004} {\bibfield  {journal} {\bibinfo  {journal} {Phys. Rev. D}\ }\textbf {\bibinfo {volume} {106}},\ \bibinfo {pages} {043004} (\bibinfo {year} {2022})},\ \bibinfo {note} {[Erratum: Phys.Rev.D 106, 109901 (2022)]},\ \Eprint {https://arxiv.org/abs/2205.03395} {arXiv:2205.03395 [hep-ph]} \BibitemShut {NoStop}%
\bibitem [{\citenamefont {Kahn}\ \emph {et~al.}(2021)\citenamefont {Kahn}, \citenamefont {Krnjaic},\ and\ \citenamefont {Mandava}}]{Kahn:2020fef}%
  \BibitemOpen
  \bibfield  {author} {\bibinfo {author} {\bibfnamefont {Y.}~\bibnamefont {Kahn}}, \bibinfo {author} {\bibfnamefont {G.}~\bibnamefont {Krnjaic}},\ and\ \bibinfo {author} {\bibfnamefont {B.}~\bibnamefont {Mandava}},\ }\bibfield  {title} {\bibinfo {title} {{Dark Matter Detection with Bound Nuclear Targets: The Poisson Phonon Tail}},\ }\href {https://doi.org/10.1103/PhysRevLett.127.081804} {\bibfield  {journal} {\bibinfo  {journal} {Phys. Rev. Lett.}\ }\textbf {\bibinfo {volume} {127}},\ \bibinfo {pages} {081804} (\bibinfo {year} {2021})},\ \Eprint {https://arxiv.org/abs/2011.09477} {arXiv:2011.09477 [hep-ph]} \BibitemShut {NoStop}%
\bibitem [{\citenamefont {Kurinsky}\ \emph {et~al.}(2020)\citenamefont {Kurinsky}, \citenamefont {Baxter}, \citenamefont {Kahn},\ and\ \citenamefont {Krnjaic}}]{Kurinsky:2020dpb}%
  \BibitemOpen
  \bibfield  {author} {\bibinfo {author} {\bibfnamefont {N.}~\bibnamefont {Kurinsky}}, \bibinfo {author} {\bibfnamefont {D.}~\bibnamefont {Baxter}}, \bibinfo {author} {\bibfnamefont {Y.}~\bibnamefont {Kahn}},\ and\ \bibinfo {author} {\bibfnamefont {G.}~\bibnamefont {Krnjaic}},\ }\bibfield  {title} {\bibinfo {title} {{Dark matter interpretation of excesses in multiple direct detection experiments}},\ }\href {https://doi.org/10.1103/PhysRevD.102.015017} {\bibfield  {journal} {\bibinfo  {journal} {Phys. Rev. D}\ }\textbf {\bibinfo {volume} {102}},\ \bibinfo {pages} {015017} (\bibinfo {year} {2020})},\ \Eprint {https://arxiv.org/abs/2002.06937} {arXiv:2002.06937 [hep-ph]} \BibitemShut {NoStop}%
\bibitem [{\citenamefont {Trickle}\ \emph {et~al.}(2020)\citenamefont {Trickle}, \citenamefont {Zhang}, \citenamefont {Zurek}, \citenamefont {Inzani},\ and\ \citenamefont {Griffin}}]{Trickle:2019nya}%
  \BibitemOpen
  \bibfield  {author} {\bibinfo {author} {\bibfnamefont {T.}~\bibnamefont {Trickle}}, \bibinfo {author} {\bibfnamefont {Z.}~\bibnamefont {Zhang}}, \bibinfo {author} {\bibfnamefont {K.~M.}\ \bibnamefont {Zurek}}, \bibinfo {author} {\bibfnamefont {K.}~\bibnamefont {Inzani}},\ and\ \bibinfo {author} {\bibfnamefont {S.~M.}\ \bibnamefont {Griffin}},\ }\bibfield  {title} {\bibinfo {title} {{Multi-Channel Direct Detection of Light Dark Matter: Theoretical Framework}},\ }\href {https://doi.org/10.1007/JHEP03(2020)036} {\bibfield  {journal} {\bibinfo  {journal} {JHEP}\ }\textbf {\bibinfo {volume} {03}},\ \bibinfo {pages} {036}},\ \Eprint {https://arxiv.org/abs/1910.08092} {arXiv:1910.08092 [hep-ph]} \BibitemShut {NoStop}%
\bibitem [{\citenamefont {Atzori~Corona}\ \emph {et~al.}(2023{\natexlab{b}})\citenamefont {Atzori~Corona}, \citenamefont {Bonivento}, \citenamefont {Cadeddu}, \citenamefont {Cargioli},\ and\ \citenamefont {Dordei}}]{AtzoriCorona:2022jeb}%
  \BibitemOpen
  \bibfield  {author} {\bibinfo {author} {\bibfnamefont {M.}~\bibnamefont {Atzori~Corona}}, \bibinfo {author} {\bibfnamefont {W.~M.}\ \bibnamefont {Bonivento}}, \bibinfo {author} {\bibfnamefont {M.}~\bibnamefont {Cadeddu}}, \bibinfo {author} {\bibfnamefont {N.}~\bibnamefont {Cargioli}},\ and\ \bibinfo {author} {\bibfnamefont {F.}~\bibnamefont {Dordei}},\ }\bibfield  {title} {\bibinfo {title} {{New constraint on neutrino magnetic moment and neutrino millicharge from LUX-ZEPLIN dark matter search results}},\ }\href {https://doi.org/10.1103/PhysRevD.107.053001} {\bibfield  {journal} {\bibinfo  {journal} {Phys. Rev. D}\ }\textbf {\bibinfo {volume} {107}},\ \bibinfo {pages} {053001} (\bibinfo {year} {2023}{\natexlab{b}})},\ \Eprint {https://arxiv.org/abs/2207.05036} {arXiv:2207.05036 [hep-ph]} \BibitemShut {NoStop}%
\end{thebibliography}%

\end{document}